**Doped-$MoSe_2$ nanoflakes/3*d* metal oxide-hydr(oxy)oxides hybrid catalysts for pH-universal electrochemical hydrogen evolution reaction**

*Leyla Najafi, Sebastiano Bellani, Reinier Oropesa-Nuñez, Alberto Ansaldo, Mirko Prato, Antonio Esau Del Rio Castillo and Francesco Bonaccorso**

Dr. Leyla Najafi, Dr. Sebastiano Bellani, Dr. Reinier Oropesa-Nuñez, Dr. Alberto Ansaldo, Dr. Antonio Esau Del Rio Castillo, Dr. Francesco Bonaccorso
Graphene Labs, Istituto Italiano di Tecnologia, via Morego 30, 16163 Genova, Italy.
Email: francesco.bonaccorso@iit.it
Dr. Reinier Oropesa-Nuñez, Dr. Francesco Bonaccorso
BeDimensional Srl, via Albisola 121, 16163, Genova, Italy.
Mirko Prato
Materials Characterization Facility, Istituto Italiano di Tecnologia, via Morego 30, 16163 Genova, Italy.

Clean hydrogen production through efficient and cost-effective electrochemical water splitting is highly promising to meeting future global energy demands. The design of Earth-abundant materials with both high activity for hydrogen evolution reaction (HER) and electrochemical stability in both acidic and alkaline environments summarize the outcomes needed for practical applications. Here, we report a non-noble 3d metal Cl-chemical doping of liquid phase exfoliated single/few-layer flakes of $MoSe_2$ for creating $MoSe_2$ nanoflakes/3d metal oxide-hydr(oxy)oxide hybrid HER-catalysts. We propose that the electron-transfer from MoSe2 nanoflakes to metal cations and the chlorine complexation-induced both neutralization, as well as the in situ formation of metal oxide-hydr(oxy)oxides on $MoSe_2$ nanoflake's surface, tailor the proton affinity of the derived catalysts, increasing the number and HER-kinetic of their active sites in both acidic and alkaline electrolytes. The electrochemical coupling between the doped-$MoSe_2$ nanoflakes/metal oxide-hydr(oxy)oxide hybrids and single-walled carbon nanotubes heterostructures further accelerates the HER process. Lastly, monolithic stacking of multiple heterostructures is reported as a facile electrode assembly strategy to achieve overpotential for a cathodic current density of $10 mAcm^{-2}$ of 0.081V and 0.064V in 0.5M $H_2SO_4$ and 1M KOH, respectively. This opens up new opportunities to address the current density vs. overpotential requirements targeted in pH-universal $H_2$ production.

## 1. Introduction

Two dimensional (2D)-transition metal dichalcogenides (TMDs), whose crystal structure is composed by covalently bonded X-M-X (M = transition metal; X = chalcogen) layers held together by van der Waals forces,[1,2,3] have been reported as noble, metal-free, highly-efficient and stable electrocatalysts for electrochemical hydrogen evolution reaction (HER).[4,5,6, 7,8,9] Theoretical[10,11,12,13] and experimental[14,15,16,17] investigations have demonstrated that the unsaturated X-edges in the natural semiconducting trigonal prismatic 2H phase of TMDs are HER electrocatalytically active in acidic media, with a Gibbs free energy of adsorbed atomic H ($\Delta G_{Hads}{}^{0}$) close to zero.[10,12] Consequently, extensive efforts have been dedicated to increasing the number of exposed active sites, including the design of nanostructured films (*e.g.* double-gyroid mesoporous[16] and vertically aligned nanostructures[18,19]), the engineering of the stoichiometry of the catalytically inert basal plane by defective[20,21] and doping treatments,[8,9,22,23,24] as well as the developments of hybrid materials (*e.g.* $CoS_x$/$MoS_x$,[25] $MoS_2$/$Mo_2C$[26] and graphene/ or carbon nanotubes (CNTs)/TMDs[27,28,29] and Ni/TMDs[30,31]). Despite recent progress, there are still experimental issues that need to be resolved before TMDs can be considered an economical material platform for the production of $H_2$, including: 1) the scalability of the synthesis of high-electrocatalytically active electrocatalysts; 2) the potential cost/benefit of the hybridization strategies; 3) the feasibility of the electrode cell assembly and 4) the HER-activity and stability in both acidic and alkaline electrolytes. In fact, it is still challenging to design scalable and cost-effective pH-universal TMD-based electrocatalysts[23,32] that are capable of competing with cathode materials found in current large-scale $H_2$ production technologies[33,34] such as Ni alloys[35,36] or high surface area noble metal coated-Ni[37] for chloro-alkaline or alkaline zero gap water[38] electrolysis units, and Pt nanoparticles supported on carbon black (Pt/C) for proton exchange membrane (PEM) electrolysis.[39,40,41]

Herein, we report the design of pH-universal efficient HER-electrocatalysts based on doped-$MoSe_2$ nanoflakes/3d metal oxide-hydr(oxy)oxide hybrids by the cost-effective material production and electrode manufacturing through solution-processed methods in an environmentally friendly alcohol-based solvent. Our electrodes achieved low $\eta_{10}$ of 0.081 V and 0.064 V in 0.5 M $H_2SO_4$ and 1 M KOH, respectively, as well as promising electrochemical stability under HER-operation, fulfilling the key-requirements for practical applications.

## 2. Results and discussion

### 2.1. Production and characterization of materials

Amongst the TMDs, we focused our attention on $MoSe_2$ because it has a higher intrinsic electrical conductivity and a lower $\Delta G_H^0$ at the edge sites than other TMDs.[28,42,43] $MoSe_2$ nanoflakes were produced by a cost-effective liquid-phase exfoliation (LPE)[44,45,46] of the bulk counterpart in 2-Propanol (IPA), followed by a sedimentation-based separation[47,48,49] (see Experimental, Synthesis of materials for additional details). **Figure 1**a,b shows the scanning electron microscopy (SEM) and atomic force microscopy (AFM) images of the as-exfoliated $MoSe_2$ (ex-$MoSe_2$), whose morphology resembles a crumpled paper-like structure.

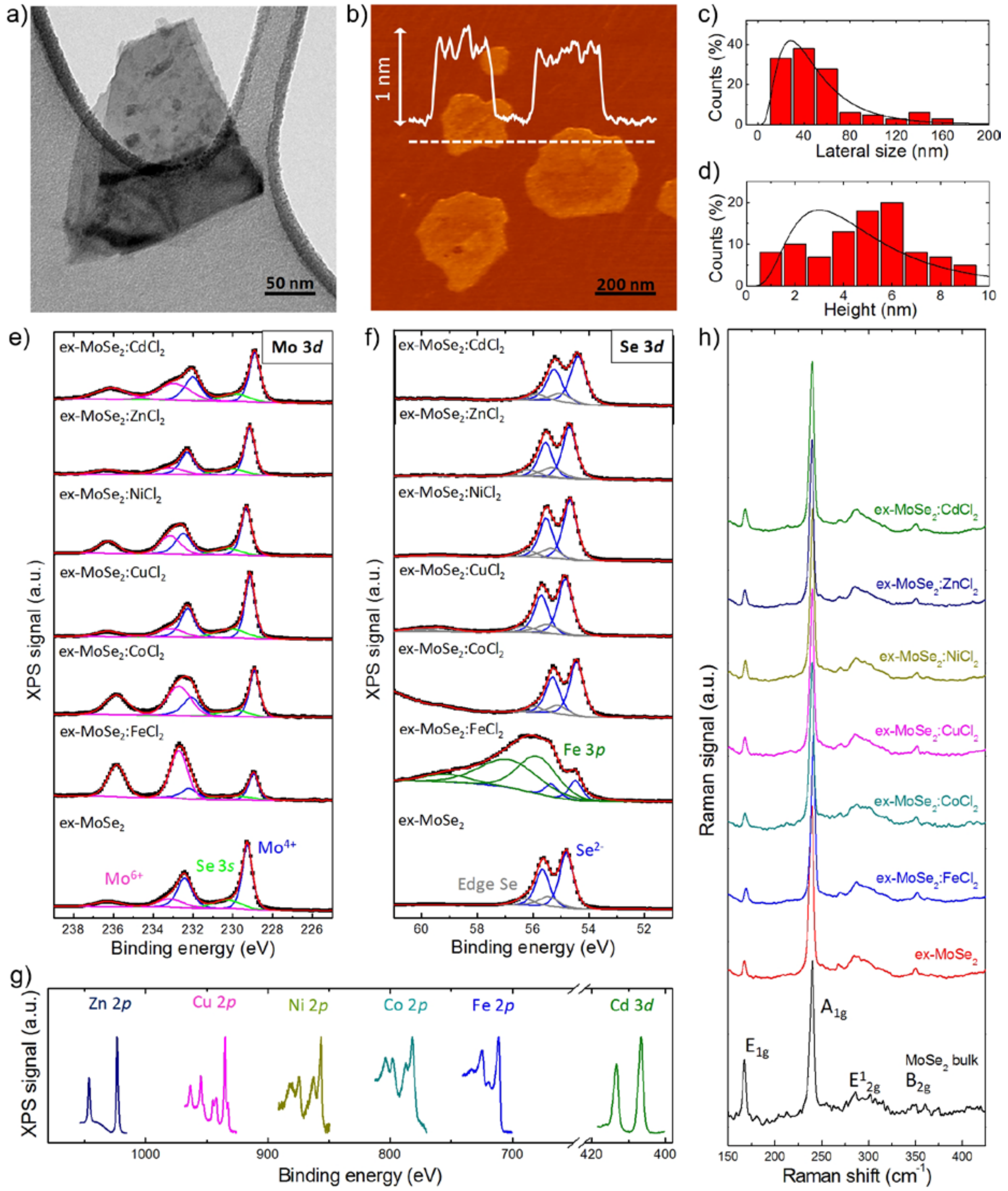

**Figure 1. Morphological, chemical and structural characterizations of pristine and $MCl_2$-doped ex-$MoSe_2$.** a) TEM image of the ex-$MoSe_2$. b) AFM image of ex-$MoSe_2$ deposited onto a mica sheet. The height profiles of two representative flakes are also shown (white line). c) TEM statistical analysis of the lateral dimension of the ex-$MoSe_2$ (calculated on 80 flakes). d) AFM statistical analysis of the thickness of the ex-$MoSe_2$ (calculated on 80 flakes from different AFM images). e,f) Mo 3*d* and Se 3*d* XPS spectra of ex-$MoSe_2$ and ex-$MoSe_2$:$MCl_2$. Their deconvolutions are also shown, evidencing the band ascribed to: $Mo^{4+}$ and $Se^{2+}$ ($MoSe_2$) (blue curves); Se 3*s* band (red curve), overlapping the Mo 3*d* XPS spectrum; $Mo^{6+}$ ($MoO_3$) (magenta curves); $Se^0$ (edge Se) (grey curves); Fe 3*p* band (olive curve), overlapping the Se 3*s* XPS spectrum. g) Fe 2*p* (blue curve), Co 2*p* (cyan curve), Cu 2*p* (magenta curve), Ni 2*p* (dark yellow curve), Zn 2*p* (Navy curve), Cd 3*d* (olive curve). h) Raman spectra of $MoSe_2$ bulk (black curve), ex-$MoSe_2$ (red curve), ex-$MoSe_2$:$FeCl_2$ (blue

curve), ex-$MoSe_2$:$CoCl_2$ (cyan curve), ex-$MoSe_2CuCl_2$ (magenta curve), ex-$MoSe_2$:$NiCl_2$ (dark yellow curve), ex-$MoSe_2$:$ZnCl_2$ (Navy curve) and ex-$MoSe_2$:$CdCl_2$ (olive curve), as-deposited on Si/$SiO_2$ substrates. The main peaks, *i.e.* the in-plane modes $E_{1g}$, $E^1_{2g}$, and $E^2_{2g}$, the out-of-plane mode $A_{1g}$ and the breathing mode $B^1_{2g}$ are named in the graph.

Statistical analysis indicates that the ex-$MoSe_2$ have a lateral size in the range of 10-170 nm (lognormal distribution peaked at ~29 nm) (Figure 1c) and a thickness up to almost 10 nm (log normal distribution peaked at ~3 nm) (Figure 1d). Thus, single/few layer ex-$MoSe_2$ were effectively produced ($MoSe_2$ monolayer thickness has been previously measured between 0.6 nm and 1 nm[50,51]). The ex-$MoSe_2$ flakes were subsequently doped through non-noble 3*d* metal (Fe, Co, Ni, Cu, Zn, Cd) chloride ($MCl_2$)-chemical doping.[52,53,54] Experimentally, the $MCl_2$-chemical doping of the ex-$MoSe_2$ was carried out by mixing an ex-$MoSe_2$ dispersion with a $MCl_2$ solution in anhydrous IPA, in which the $MCl_2$ dissociate in $M^{2+}$ and $2Cl^-$ (being the $MCl_2$ solution concentration (0.4 g $L^{-1}$) inferior to the solubility limit of $MCl_2$ in alcohol, >> 1 g $L^{-1}$),[55,56] thus obtaining ex-$MoSe_2$:$MCl_2$ dispersions in IPA (1:1 molar ratio). These underwent an ultrasonication treatment, during which we suppose a doping process occurred via "cascade reaction" as follows:

*step 1*:

a) ex-$MoSe_2 + M^{2+} + 2Cl^- \rightleftarrows$ ex-$MoSe_2^{2+} + M^0\downarrow + 2Cl^-$ and/or

b) ex-$MoSe_2 + 2M^{2+} + 4Cl^- \rightleftarrows$ ex-$MoSe_2^{2+} + M^0\downarrow + MCl_4^{2-}$

*step 2:*

a) ex-$MoSe_2^{2+} + 2Cl^- \rightleftarrows$ ex-$MoSe_2$–2Cl and/or

b) ex-$MoSe^{2+} + MCl_4^{2-} \rightleftarrows$ ex-$MoSe_2$–$MCl_4$

The doping mechanism is initiated by an electron transfer from ex-$MoSe_2$ to $M^{2+}$ (*step 1*). This is caused by the high electronegativity of the latter, which has been theoretically estimated in an aqueous solution-phase as 2.636, 2.706, 2.891, 2.952, 2.796 and 2.660 Pauling units for $Fe^{2+}$, $Co^{2+}$, $Ni^{2+}$, $Cu^{2+}$, $Zn^{2+}$ and $Cd^{2+}$, respectively.[57] These values agree with the Irving-Williams order of transition metal complexes[58] and they have to be considered as underestimations in comparison with those that correspond to an IPA-solution because of the higher polarity of the $H_2O$ (~10.2)[59], whose solvation weakens the electron-accepting power of cations[57,60] than that of IPA (~3.9)[59]. A similar doping initiation has been proposed for $MCl_2$- and $MCl_3$-doped graphene[61,62] and carbon nanotubes (CNTs).[63] Lastly, a neutralization of the charged species ($MoSe_2^{2+}$, $MCl_4^{2-}$ and $Cl^-$) occurs by creating ex-$MoSe_2$–2Cl and ex-$MoSe_2$–$MCl_4$ complexes (*step 2*). Here, the electron cloud shifts toward the Cl-based centers, creating a net charge displacement, leading to a p-type doping of the ex-$MoSe_2$.[61,62,63]] Notably, the electron depletion of $MCl_2$-doped TMD nanoflakes has been previously exploited for designing p-doped $MoS_2$-based hole transport layers in organic solar cells[64] and $H_2$-evolving photocathodes.[65] The chemical and electronic state within the exposed surfaces of $MCl_2$-doped ex-$MoSe_2$ (ex-$MoSe_2$:$MCl_2$) were evaluated by x-ray photoelectron spectroscopy (XPS) analysis (Figure 1e-g). Figure 1e,f show the Mo 3*d* and Se 3*d* XPS spectra of ex-$MoSe_2$:$MCl_2$ (except for M = Cu). These data evidence a uniform shift (in the range of 0.1-0.4 eV) towards a lower binding energy compared to those of the as-produced ex-$MoSe_2$. This indicates that the amount of energy that is required to remove an electron from the ex-$MoSe_2$:$MCl_2$ surface is lower than that needed for the ex-$MoSe_2$ that means that the electronic density in the ex-$MoSe_2$ surface increases after $MCl_2$-doping. Actually, the downshift also agrees with the decrease of the Fermi level upon the p-type $MCl_2$-doping of ex-$MoSe_2$ and the upward band-bending appearance localized at the doped/undoped regions' interface.[52,66,67] In the Mo 3*d* spectra (Figure 1e), the two peaks located at ~229 eV and ~232 eV are attributed to the Mo $3d_{5/2}$ and Mo $3d_{3/2}$ peaks of the $Mo^{4+}$ state in $MoSe_2$.[68,69] The distinct binding energy peaks associated with the M 2*p* or 3*d* doublets (Figure 1g) are ascribed to the oxidized states of M (*i.e.* $Fe^{3+}$, $Co^{2+}$, $Ni^{2+}$, $Cu^{1+}$ or $Cu^{2+}$, $Zn^{2+}$ and $Cd^{2+}$), corresponding to both $MCl_2$ residuals and metal oxide (*e.g.* MO for $M^{2+}$, $M_2O_3$ for $M^{3+}$) or metal hydr(oxy)oxides (*e.g.* $M(OH)_2$ for $M^{2+}$ or $M(OH)_3$ and M(OH)O for $M^{3+}$) originating from the reaction between $M^0\downarrow$ and $O_2$ or $H_2O$, after exposure to ambient.

Additional details of the XPS analysis are reported in the Supplementary Information (ESI). In short, XPS data show that metal oxide/hydr(oxy)oxides have been formed *in situ* on the surface of the ex-$MoSe_2$, creating doped-$MoSe_2$ nanoflakes/3*d* metal oxide-hydr(oxy)oxides hybrids. The structural topological properties of the ex-$MoSe_2$ and ex-$MoSe_2$:$MCl_2$, as well as those of their $MoSe_2$ bulk counterparts, were evaluated by Raman spectroscopy measurements. According to group theory analysis, $MoSe_2$ bulk is characterized by four Raman active modes: three in-plane $E_{1g}$, $E^1_{2g}$, and $E^2_{2g}$, and one out-of-plane $A_{1g}$ of $D_{6h}$ point group symmetry.[70,71,72] After exfoliation, the interlayer vibrational mode $B_{2g}$ is activated by the breakdown of the translational symmetry that occurs in few-layer $MoSe_2$ flakes.[73] Representative spectra of $MoSe_2$ bulk, ex-$MoSe_2$ and ex-$MoSe_2$:$MCl_2$ are reported in Figure 1h. The $A_{1g}$ mode is located at ~241 $cm^{-1}$ for the $MoSe_2$ bulk, while it red-shifts to ~239 $cm^{-1}$ for the ex-$MoSe_2$, which is in agreement with the softening of the vibrational mode that is accompanied by a reduction in flake thickness.[74,75,76,77,78] The in-plane $E^1_{2g}$ mode is observed at ~287 $cm^{-1}$ for both samples.[74-76] The intensity ratio between the $A_{1g}$ and $E^1_{2g}$ modes ($I(A_{1g})/I(E^1_{2g})$) (~21) and the presence of the $B_{2g}$ mode (located at ~352 $cm^{-1}$) in the ex-$MoSe_2$ spectra indicate the few-layer flake structure.[75,79,80] The activation of the mode $E_{1g}$ is due to a resonance-induced symmetry breaking effect.[71,81] Moreover, the energy of this mode, being independent of the number of layers,[82] does not change between the $MoSe_2$ bulk and ex-$MoSe_2$. The Raman statistical analysis is reported in ESI (**Figure S1**). Although XPS analysis evidences surface modifications of ex-$MoSe_2$ after $MCl_2$-doping, the presence of the Raman peaks that are attributed to ex-$MoSe_2$ indicates that the crystal structure of undoped flakes is preserved. Therefore, doping-induced chemical modifications are excluded, suggesting that molecular van der Waals complexes have formed between ex-$MoSe_2^{2+}$ and $Cl^-$ or $MCl_4^{2-}$. These complexes originated from the charge neutralization through physisorption and/or chemisorption of the ionic species to the ex-$MoSe_2$ during the $MCl_2$-doping process (*step 2*).[53,83,84]

## 2.2. Production and electrochemical characterization of electrocatalysts

The HER-activities of ex-$MoSe_2$ and ex-$MoSe_2$:$MCl_2$, which were deposited on glassy carbon substrates (ex-$MoSe_2$ mass loading: 0.2 mg $cm^{-2}$), were evaluated both in acidic (0.5 M $H_2SO_4$) and alkaline (1 M KOH) solutions **(Figure 2)**.

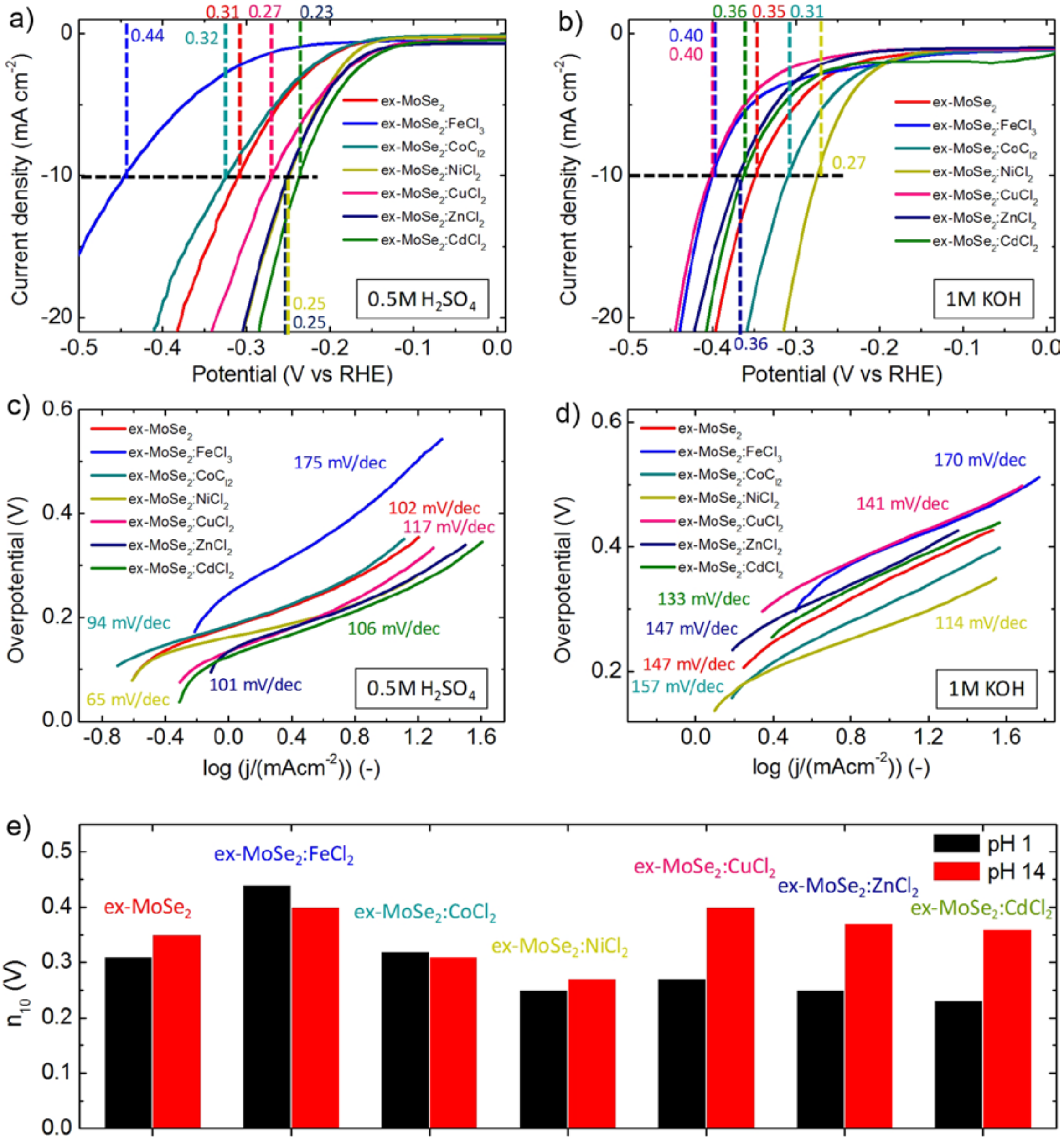

**Figure 2. Electrochemical characterization of the HER-activity of ex-$MoSe_2$ and ex-$MoSe_2$:$MCl_2$ in acidic and alkaline solutions.** a,b) LSV curves for ex-$MoSe_2$ and ex-$MoSe_2$:$MCl_2$ in 0.5 M $H_2SO_4$ (pH 1) and 1 M KOH (pH 14), recorded during initial potential sweeps. c,d) Corresponding Tafel plots of the LSV curves in (a) and (b). e) Comparison between the overpotential at 10 mA $cm^{-2}$ ($n_{10}$) of cathodic currents of ex-$MoSe_2$ and ex-$MoSe_2$:$MCl_2$ in acidic (black histograms) and alkaline solutions (red histograms).

Theoretically, the HER in acidic solution proceeds with an initial discharge of the hydronium ion ($H_3O^+$) and the formation of intermediates *i.e.* atomic H adsorbed on the electrocatalyst surface ($H_{ads}$), in the so-called Volmer step ($H_3O^+ + e^- \rightleftarrows H_{ads} + H_2O$), followed by either an electrochemical Heyrovsky step ($H_{ads} + H_3O^+ + e^- \rightleftarrows H_2 + H_2O$) or a chemical Tafel recombination step ($2H_{ads} \rightleftarrows H_2$). In alkaline media, the $H_{ads}$ is formed by discharging $H_2O$ ($H_2O + e^- \rightleftarrows H_{ads} + OH^-$). Then, either a Heyrovsky step ($H_2O + H_{ads} + e^- \rightleftarrows H_2 + OH^-$) or a chemical Tafel recombination step ($2H_{ads} \rightleftarrows H_2$) occurs. The overpotential at a cathodic current density of 10 mA $cm^{-2}$ ($\eta_{10}$), the Tafel slope and the exchange current density ($j_0$) are the Figures of Merit (FoM) that are needed to assess the HER-activity.[85,86] The last two of these FoM can be extracted from the linear portion of the Tafel plots, which show the relation between the overpotential and the current density of the electrodes, and are in agreement with the Tafel equation (see the Experimental, Characterization of electrodes). The Tafel slope measures the potential increase that is required to improve the current density by 1 order of magnitude. Fundamentally, it is used to evaluate the HER mechanism at the electrode/electrolyte interface.[85,86] For an insufficient $H_{ads}$ surface coverage, the Volmer reaction is the rate limiting step of the HER, and a theoretical Tafel slope of 120 mV $dec^{-1}$ is observed. On the other hand, for a borderline case in which there is a high $H_{ads}$ surface coverage (*i.e.* in which the $\Delta G_{Hads}{}^0$ is close to zero), the HER-kinetic is dominated by the Heyrovsky or Tafel reaction, and a Tafel slope of 40 or 30 mV $dec^{-1}$ is detected. The $j_0$ is linked to the number of available HER-active sites and their HER-kinetics.[85,86] Figure 2a,b show linear sweep voltammetry (LSV) curves of ex-$MoSe_2$ and ex-$MoSe_2$:$MCl_2$, in acidic and alkaline solutions, respectively, while Figure 2c,d report the corresponding Tafel plots. The extrapolated $\eta_{10}$ are reported in Figure 2e. The LSV data show that the HER-activity of ex-$MoSe_2$ was affected by the $MCl_2$-doping.

In the acidic solution, the ex-$MoSe_2$ doped with $CdCl_2$, $ZnCl_2$, $CuCl_2$ and $NiCl_2$ showed higher cathodic current densities than the pristine ex-$MoSe_2$, while the $CoCl_2$- and $FeCl_2$- doping decreased the HER-activity. Thus, the $\eta_{10}$ decreased from 0.445 V for the ex-$MoSe_2:FeCl_2$ to 0.235 V for the most active case of ex-$MoSe_2:CdCl_2$. The Tafel slope of ex-$MoSe_2:NiCl_2$ significantly decreased from 0.102 V $dec^{-1}$ in the ex-$MoSe_2$ to 0.065 V $dec^{-1}$, while it was marginally affected by $CoCl_2$-, $CuCl_2$-, $ZnCl_2$- and $CdCl_2$- doping (Figure 2c,d). The $j_0$ was significant affected by $NiCl_2$-doping, but there was no remarkable variation for the others $MCl_2$-dopings (**Figure S2**). In the alkaline solution, $CoCl_2$ and $NiCl_2$ were the most effective dopants for increasing the HER-activity of the ex-$MoSe_2$. The corresponding $\eta_{10}$ were 0.308 V and 0.273 V, respectively, which were lower than that of the ex-$MoSe_2$ (0.349 V). $CdCl_2$ and $ZnCl_2$ dopants did not significantly affect the ex-$MoSe_2$ HER-activity, which decreased for both ex-$MoSe_2:CuCl_2$ and ex-$MoSe_2:FeCl_2$. As was the case in the acidic media, $NiCl_2$ was the most effective dopant for decreasing the Tafel slope of the ex-$MoSe_2$ (from 0.147 to 0.114 V $dec^{-1}$). The ex-$MoSe_2$ and the ex-$MoSe_2:MCl_2$ displayed comparable $j_0$ values (in the range of 10-30 $\mu A\ cm^{-2}$), with the only exception of ex-$MoSe_2:CuCl_2$, which showed a higher $j_0$ value (around 70 $\mu A\ cm^{-2}$). These results indicate that $MCl_2$-doping can be effective in enhancing the HER-activity of ex-$MoSe_2$ in both acidic and alkaline solutions. Once the HER-activity of the ex-$MoSe_2:MCl_2$ had been established, flexible, self-standing heterostructures between single-walled carbon nanotubes (SWCNTs) and ex-$MoSe_2:MCl_2$ (SWCNTs/ex-$MoSe_2:MCl_2$), produced *via* the sequential vacuum filtration of the SWCNT and ex-$MoSe_2:MCl_2$, were investigated as flexible HER-active electrodes compatible with high-throughput scalable industrial manufacturing. The rationale of the choice of these heterostructures was based on our recent works,[87,88] in which we demonstrated a long-range ($\geq$ 1 $\mu m$) electrochemical coupling between TMDs and graphene or SWCNT paper (substrates)[89,90] for increasing the HER-activity of TMDs, without resorting the synthesis of hybrid TMDs/carbon material compounds.[28-31]

**Figure 3**a,b report the top-view SEM images of the SWCNT paper (SWCNT mass loading: 0.64 mg $cm^{-2}$) and SWCNTs/ex-$MoSe_2$ (SWCTNs mass loading: 0.64 mg $cm^{-2}$; ex-$MoSe_2$ mass loading: 0.64 mg $cm^{-2}$).

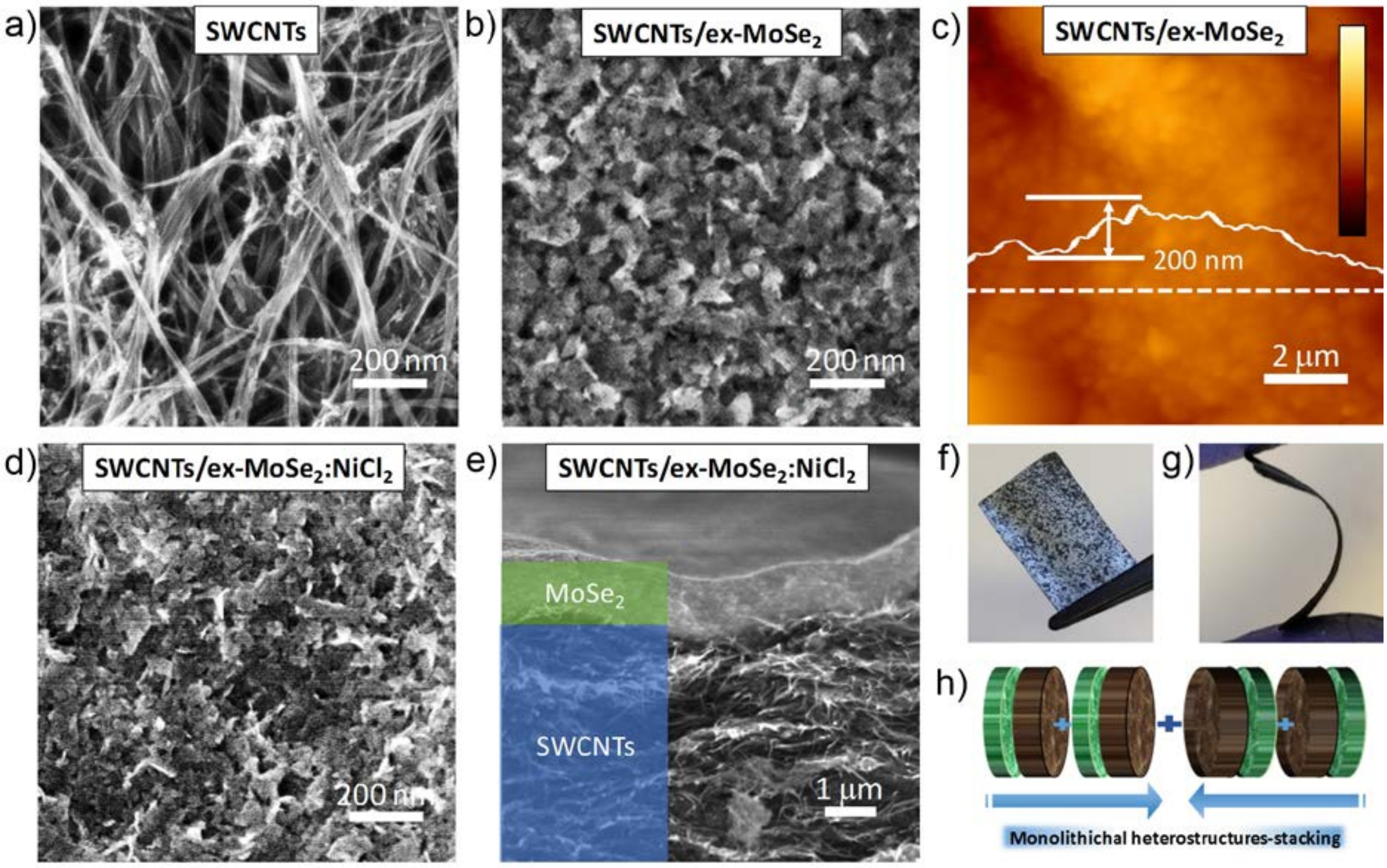

**Figure 3. Solution-processed, flexible and self-standing heterostructured electrodes between SWCNTs and SWCNTs/ex-$MoSe_2$:$MCl_2$ prepared via sequential vacuum filtration of the material dispersions.** a,b) Top-view SEM images of SWCNTs bucky paper and SWCNTs/ex-$MoSe_2$:$NiCl_2$. c) AFM images SWCNTs/ex-$MoSe_2$. Height profile along representative cross section (white dashed lines) is also shown. The z-scale bar is 1 µm. d,e) Top-view and cross-sectional SEM images of SWCNTs/ex-$MoSe_2$:$NiCl_2$. f,g)Top-view and side view photographs of SWCNTs/ex-$MoSe_2$:$NiCl_2$ (electrode area: 0.8x1.5 $cm^2$. In panel g) the electrode was manually bended in order to show its mechanical flexibility. H) Sketch of the monolithical heterostructures-stacking (up to 4-stacked heterostructures).

The surface of the SWCNT paper consists of a mesoporous network with a bundle-like arrangement (Figure 3a). The surface of the SWCNT paper is clearly modified by the ex-$MoSe_2$ overlay (Figure 3b), which is characterized by the flaked nature of the ex-$MoSe_2$ (see TEM and AFM analysis, Figure 1). The AFM images of the electrode surfaces (**Figure S3** and Figure 3c) evidence morphologies similar to those observed by SEM. The calculated roughness average (Ra) values is ~103 nm and ~70 nm for SWCNT paper and SWCNTs/ex-$MoSe_2$ surfaces, respectively. These values indicate that the ex-$MoSe_2$ deposition flattens the SWCNT paper surface. The top-view SEM image of a representative SWCNTs/ex-$MoSe_2$:$NiCl_2$ (Figure 3d) does not evidence any significant surface changes in comparison to that of SWCNT/ex-$MoSe_2$. The cross-sectional SEM images of SWCNTs/ex-$MoSe_2$:$NiCl_2$ (Figure 3e) shows a bilayer-like architecture. The thickness of the ex-$MoSe_2$:$NiCl_2$ overlay is ~1 μm. The elemental energy-dispersive X-ray spectroscopy (EDX) analysis of a top view SEM image of SWCNTs/ex-$MoSe_2$:$NiCl_2$ (**Figure S4**) shows the uniform distribution of Ni onto the heterostructure's surface, thus indicating that the $NiCl_2$-doping of ex-$MoSe_2$ is uniform. Interestingly, the EDX analysis of a cross-sectional SEM image of SWCNTs/ex-$MoSe_2$:$NiCl_2$ (**Figure S5**) reveals that Ni is also distributed uniformly along the vertical direction of the heterostructure, including the SWCNT paper. This means that, during deposition of the ex-$MoSe_2$:$MCl_2$, the elemental metal (*i.e.* $M^0$↓) and, eventually, the $MCl_2$ residuals also infiltrated the SWCNT paper. Thus, the formation of metal oxides (MO or $M_2O_3$) or metal hydr(oxy)oxides ($M(OH)_2$, $M(OH)_3$ or M(OH)O) on SWCNT paper and the p-doping of the SWCNT[63] can also occur during the fabrication of the heterostructures. It is worth noting that the p-doping of the SWCNT can enhance the electron transfer from the SWCNT papers to the active sites of ex-$MoSe_2$:$MCl_2$, as experimentally observed on a highly-sensitive, p-n heterojunction diode-photodetector with a fast response (< 15 μs) and tuneable gate using SWCNTs and single-layer $MoS_2$ as p-type and n-type semiconductors.[91] Figure 3f,g show top-view and side-view photographs of SWCNTs/ex-$MoSe_2$:$NiCl_2$. The picture in Figure 3g was taken on manually bended SWCNTs/ex-$MoSe_2$:$NiCl_2$ in order to illustrate its mechanical flexibility, which can enable versatile designs to be exploit for advanced solar fuel devices including flexible photoelectrochemical cells[92,93] and $H_2$ storage systems[93,94,95]. Actually, the designed heterostructures also offer interconnected pores which improve the transportation of evolved $H_2$,[96,97] enabling multiple heterostructure to be stacked in a single electrode. Therefore, 2- and 4-stacked heterostructures were then assembled (Figure 3h). The monolithical stacking potentially allow the electrodes to fulfil the areal performance requirements of energy renewable buffer units (cathodic current density > 100 mA $cm^{-2}$ at an overpotential less than 200 mV).[33-41] It is worth noting that, by selecting the number of heterostructures, the mass loading of the active materials is no longer limited as is the case with the single heterostructures, in which the fragmentation of films with a mass loading superior to a few mg $cm^{-2}$ occur during their deposition and/or HER-operation (*i.e.* gas evolution).[98]

**Fig 4**a,b report the *iR*-corrected LSV curves of the SWCNTs/ex-$MoSe_2$:$CdCl_2$ and SWCNTs/ex-$MoSe_2$:$NiCl_2$ in acidic and alkaline solutions, respectively, together with those of the 2- and 4-stacked corresponding heterostructures (labeled as: 2- SWCNTs/ex-$MoSe_2$:$CdCl_2$ and 2-SWCNTs/ex-$MoSe_2$:$NiCl_2$; 4-SWCNTs/ex-$MoSe_2$:$CdCl_2$, and 4-SWCNTs/ex-$MoSe_2$:$NiCl_2$). The LSV curves of SWCNTs/ex-$MoSe_2$:$NiCl_2$ and SWCNTs/ex-$MoSe_2$:$CoCl_2$ in acidic and alkaline solutions, respectively, are also reported for comparison (ex-$MoSe_2$:$NiCl_2$ and ex-$MoSe_2$:$CoCl_2$ were the second most HER-active ex-$MoSe_2$:$MCl_2$ in the aforementioned solutions). The LSV curves of commercial Pt/C are also shown as benchmark. These results show that the most HER-active hybrid heterostructures are those based on ex-$MoSe_2$:$CdCl_2$ and ex-$MoSe_2$:$NiCl_2$, in acidic and alkaline solutions, respectively, thus proving that the $MCl_2$-doping and the electrochemical coupling between SWCNTs and ex-$MoSe_2$ had a superimposed HER-assistive effect. Notably, the 4-stacked heterostructures yielded current densities higher than 100 mA $cm^{-2}$ at an overpotential lower than 0.2 V, both in acidic and alkaline media. A rigorous kinetic analysis of the HER, *i.e.* the establishment of the Tafel slope and the $j_0$, was not carried out because of the unambiguous results that derive from the presence of the SWCNT paper. In fact, SWCNTs hold a high surface area, which leads to a remarkable capacitive current density (in the order of 1 mA $cm^{-2}$) even at a low LSV sweep voltage rate (≤ 5 mV $s^{-1}$). This can be the cause of misleading interpretations of the estimated kinetic parameters.[99] Thus, the HER-activity was evaluated only by analysing the $\eta_{10}$, as extrapolated from LSV measurements (Figure 2c,d).

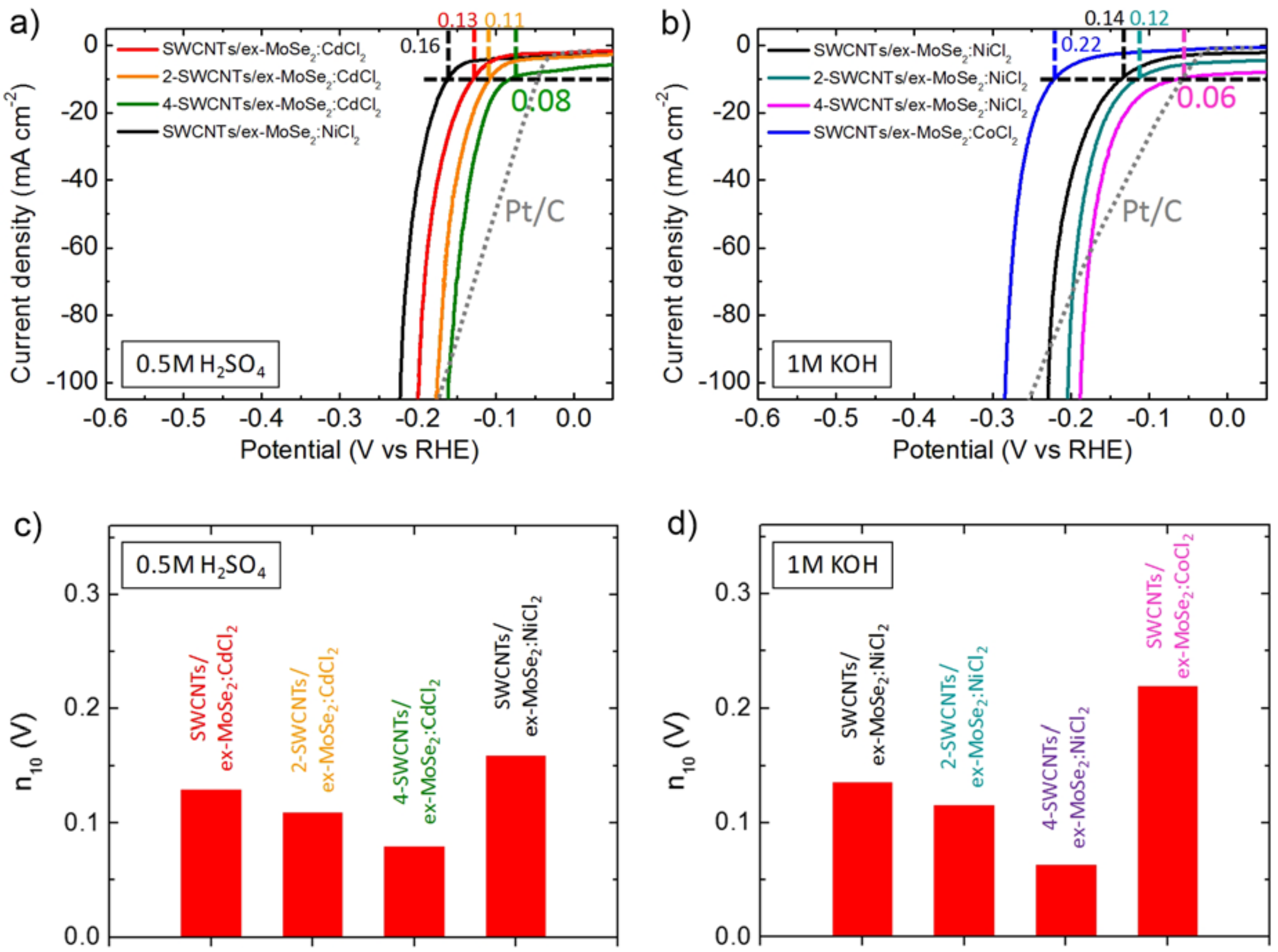

**Figure 4. Electrochemical characterization of the HER-activity of hybrid heterostructures (SWCNTs/ex-$MoSe_2$:$MCl_2$) and monolithically 2- or 4-stacked heterostructures (2-SWCNTs/ ex-$MoSe_2$:$MCl_2$ or 4-SWCNTs/ ex-$MoSe_2$:$MCl_2$).** a) LSV curves for SWCNTs/ ex-$MoSe_2$:$CdCl_2$, 2-SWCNTs/ ex-$MoSe_2$:$CdCl_2$, 4-SWCNTs/ ex-$MoSe_2$:$CdCl_2$ and SWCNTs/ex-$MoSe_2$:$NiCl_2$ in 0.5 M $H_2SO_4$ (pH 1). The LSV curve of commercial Pt/C is also shown for comparison. b) LSV curves for SWCNTs/ex-$MoSe_2$:$NiCl_2$, 2-SWCNTs/ ex-$MoSe_2$:$NiCl_2$, 4-SWCNTs/ ex-$MoSe_2$:$NiCl_2$ and SWCNTs/ex-$MoSe_2$:$CoCl_2$ in 1 M KOH (pH 14). The LSV curve of commercial Pt/C is also shown for comparison. c,d) Comparison between the $n_{10}$ corresponding to LSV curves in (a) and (b). e,f) Chronoamperometry measurements (j-t curves) at a fixed potential of -0.18 V *vs.* RHE for: SWCNTs/ex-$MoSe_2$:$CdCl_2$ and SWCNTs/ex-$MoSe_2$:$NiCl_2$ in 0.5 M $H_2SO_4$ (e); SWCNTs/ex-$MoSe_2$:$NiCl_2$ and SWCNTs/ex-$MoSe_2$:$CoCl_2$ in 1 M KOH (f). The insets of panels show the chronoamperometry measurements of ex-$MoSe_2$:$CdCl_2$ and ex-$MoSe_2$:$NiCl_2$ deposited on glassy carbon in 0.5 $H_2SO_4$ (e); ex-$MoSe_2$:$NiCl_2$ and ex-$MoSe_2$:$CoCl_2$ deposited on glassy carbon in 1 M KOH (f).

**Table 1** lists the $\eta_{10}$ values that have been measured for the different electrodes (including those using glassy carbon as a substrate, Figure 2) in both acidic and alkaline media. In the acidic medium, the $\eta_{10}$ was 0.13 V and 0.16 V for SWCNTs/ex-$MoSe_2$:$CdCl_2$ and SWCNTs/ex-$MoSe_2$:$NiCl_2$, respectively. After stacking the SWCNTs/ex-$MoSe_2$:$CdCl_2$ heterostructures, the $\eta_{10}$ decreased to 0.081 V for 4-SWCNTs/ex-$MoSe_2$:$CdCl_2$. In the alkaline medium, the $\eta_{10}$ was 0.14 V and 0.22 V for SWCNTs/ex-$MoSe_2$:$NiCl_2$ and SWCNTs/ex-$MoSe_2$:$CoCl_2$, respectively. After stacking the SWCNTs/ex-$MoSe_2$:$NiCl_2$, the $\eta_{10}$ decreased to 0.064 V for 4-SWCNTs/ex-$MoSe_2$:$NiCl_2$.

**Table 1** Summary of the overpotential at 10 mA cm$^{-2}$ ($\eta_{10}$) for the produced electrodes in both acidic and alkaline media.

| **Electrode** | **$\eta_{10}$ (V)** | |
|---|---|---|
| | **0.5 M $H_2SO_4$** | **1 M KOH** |
| ex-$MoSe_2$ | 0.308 | 0.445 |
| ex-$MoSe_2$:$FeCl_2$ | 0.445 | 0.400 |
| ex-$MoSe_2$:$CoCl_2$ | 0.323 | 0.308 |
| ex-$MoSe_2$:$NiCl_2$ | 0.248 | 0.273 |
| ex-$MoSe_2$:$CuCl_2$ | 0.269 | 0.403 |
| ex-$MoSe_2$:$ZnCl_2$ | 0.250 | 0.369 |
| ex-$MoSe_2$:$CdCl_2$ | 0.235 | 0.363 |
| SWCNTs/ex-$MoSe_2$:$CoCl_2$ | n.d. | 0.221 |
| SWCNTs/ex-$MoSe_2$:$NiCl_2$ | 0.162 | 0.136 |
| SWCNTs/ex-$MoSe_2$:$CdCl_2$ | 0.129 | n.d. |
| 2-SWCNTs/ex-$MoSe_2$:$NiCl_2$ | n.d. | 0.116 |
| 2-SWCNTs/ex-$MoSe_2$:$CdCl_2$ | 0.108 | n.d. |
| 4-SWCNTs/ex-$MoSe_2$:$NiCl_2$ | n.d. | 0.064 |
| 4-SWCNTs/ex-$MoSe_2$:$CdCl_2$ | 0.081 | n.d. |

Notably, the heterostructures' stacking, together with $MCl_2$ doping and SWCNTs/ex-$MoSe_2$:$MCl_2$ coupling, was effective in reaching $\eta_{10}$ which approach those of noble metal-based HER-catalysts in both acidic and alkaline media.[100,101,102] Actually, 4-SWCNTs/ex-MoSe2:$CdCl_2$ and 4-SWCNTs/ex-MoSe2:$NiCl_2$ exhibited lower overpotential at a cathodic current of 100 mA cm$^{-2}$ with respect to commercial Pt/C (Figure 4a,b). Although efficient TMD-based HER-electrocatalysts in acidic solutions have been reported recently in literature,[4-9,19,27-29,87,88] the corresponding HER kinetics in alkaline electrolytes typically suffers from a high overpotential ($\eta_{10}$ > 0.22 V).[23,25] Actually, the high kinetic energy barrier of the initial Volmer step and the strong adsorption of the formed $OH^-$ on the surfaces of TMDs are considered responsible for the sluggish HER kinetics in alkaline solutions.[32,103] Recently, Ni-doped $MoS_2$[8], $MoS_2$-based composites with Ni[30,31] and $CoS_2$[25], as well as well as TMD/metal hydroxide hybrids[104] have been reported as efficient HER-electrocatalysts in alkaline solutions, thus enabling prospective opportunities for using TMDs in that alkaline solutions which are used in current large-scale $H_2$ production technologies. In this context, the development of our high HER-active TMD-based electrodes with scalable and cost-effective techniques such as LPE, represent a further breakthrough with regard to creating innovative and competitive industrial technology. **Table S1** shows a comparison between the $\eta_{10}$ of our best electrocatalysts in acidic and alkaline media and those achieved by noble metal-free HER-electrocatalysts reported in literature. In particular, the measured $\eta_{10}$ of the 4-SWCNTs/ex-$MoSe_2$:$CdCl_2$ in the acidic medium (0.081 V) and the 4-SWCNTs/ex-$MoSe_2$:$NiCl_2$ in the alkaline medium (0.064 V) are inferior to those of the relevant TMD-based electrocatalysts (regardless of whether they were measured in acidic or alkaline solutions)[15, 16, 19, 21,25,27-29,77,105,106,107,108,109,110], and inferior or comparable to those of the most active noble metal-free electrocatalysts.[111,112,113,114,115,116]

In addition to HER-activity, electrocatalytic stability is also an important criterion for the commercial exploitation of electrocatalysts. **Figure 5** show the chronoamperometry measurements at a fixed potential of -0.18 V *vs*. RHE for SWCNTs/ex-$MoSe_2$:$CdCl_2$ and SWCNTs/ex-$MoSe_2$:$NiCl_2$ in the acidic medium, and SWCNTs/ex-$MoSe_2$:$NiCl_2$ and SWCNTs/ex-$MoSe_2$:$CoCl_2$ in the alkaline medium. In the acidic medium (Figure 5a) the SWCNTs/ex-$MoSe_2$:$CdCl_2$ and the SWCNTs/ex-$MoSe_2$:$NiCl_2$ slightly degraded, retaining ~86% and ~84%, respectively, of the initial current densities. This electrochemical performances reduction could be attributed to both material degradation in acidic solution and mechanical stresses generated by intense $H_2$ bubbling (corresponding to a starting current density > 200 mA $cm^{-2}$) during the measurements. Actually, ex-$MoSe_2$ has been demonstrated to be stable in 1 M $H_2SO_4$ in our previous work.[88] However, metal oxide-hydr(oxy)oxides dissolution in acid could also occur. In order to monitor the possible release of metals during chronoamperometry measurements, elemental analysis by inductively coupled plasma optical emission spectroscopy (ICP-OES) measurements were performed on our samples. Specifically, we carried out ICP-OES measurements on digested solutions prepared from the 0.5 M $H_2SO_4$ solution recovered after carrying out HER measurements of ex-$MoSe_2$:$CdCl_2$, ex-$MoSe_2$:$NiCl_2$, SWCNTs/ex-$MoSe_2$:$CdCl_2$ and SWCNTs/ex-$MoSe_2$:$NiCl_2$ (additional details on sample preparation are reported in Experimental Section). For ex-$MoSe_2$:$CdCl_2$ and ex-$MoSe_2$:$NiCl_2$, the results indicated the presence of Cd (~1.8 ppm) and Ni (~1.1 ppm), respectively. Noteworthy, for SWCNTs/ex-$MoSe_2$:$CdCl_2$ and SWCNTs/ex-$MoSe_2$:$NiCl_2$, the release of Cd decreased by ~45% (concentration of ~1.0 ppm), while Ni was not detected. This indicates that SWCNTs can suppress the physical ripening/dissolution of metal oxide-hydr(oxy)oxides), in agreement with the higher durability of SWCNTs/ex-$MoSe_2$:$CdCl_2$ and SWCNTs/ex-$MoSe_2$:$NiCl_2$ compared to that of ex-$MoSe_2$:$CdCl_2$ and ex-$MoSe_2$:$NiCl_2$ (inset to Figure 5a). In the alkaline medium (Figure 5b), the SWCNTs/ex-$MoSe_2$:$NiCl_2$ and the SWCNTs/ex-$MoSe_2$:$CoCl_2$ showed a catalytic activation, which increased the current densities by 13% during the first 50 min. After, the SWCNTs/ex-$MoSe_2$:$NiCl_2$ continued to manifest activation effects, reaching an overall increase in the current density by ~20% after 1000 min, while the SWCNTs/ex-$MoSe_2$:$CoCl_2$ slightly degraded, retaining ~99% of the initial current densities. Notably, similar trends were observed for the ex-$MoSe_2$:$MCl_2$ deposited on glassy carbon, as is illustrated in the insets in Figure 5. X-ray photoelectron spectroscopy measurements revealed no chemical degradation of ex-$MoSe_2$ during chronoamperometry measurement in alkaline solution (**Figure S6**). It is worth to mention that the hybridization of ex-$MoSe_2$ with metal oxide-hydr(oxy)oxides could also enhance the HER-durability of ex-$MoSe_2$ in alkaline condition. In fact, recent works[104,117,118,119] revealed an improvement of the stability in alkaline conditions of HER-electrocatalysts, including TMDs,[104] after their hybridization with metal hydroxides, which were used to boost the HER-activity.[25,120,121] Moreover, we speculate that the activation process observed in our electrocatalysts could be caused by a structural material reorganization determined by mechanical stresses generated by $H_2$ bubbling through the structure of the electrodes. Activation effects were significantly pronounced for ex-$MoSe_2$:$NiCl_2$ and ex-$MoSe_2$:$CoCl_2$ (inset to Figure 5b), while the mesoporous structure of SWCNTs:ex-$MoSe_2$:$NiCl_2$ and SWCNTs/ex-$MoSe_2$:$CoCl_2$ (Figure 3e) might attenuate mechanical stresses by facilitating the $H_2$ diffusion through the electrodes.

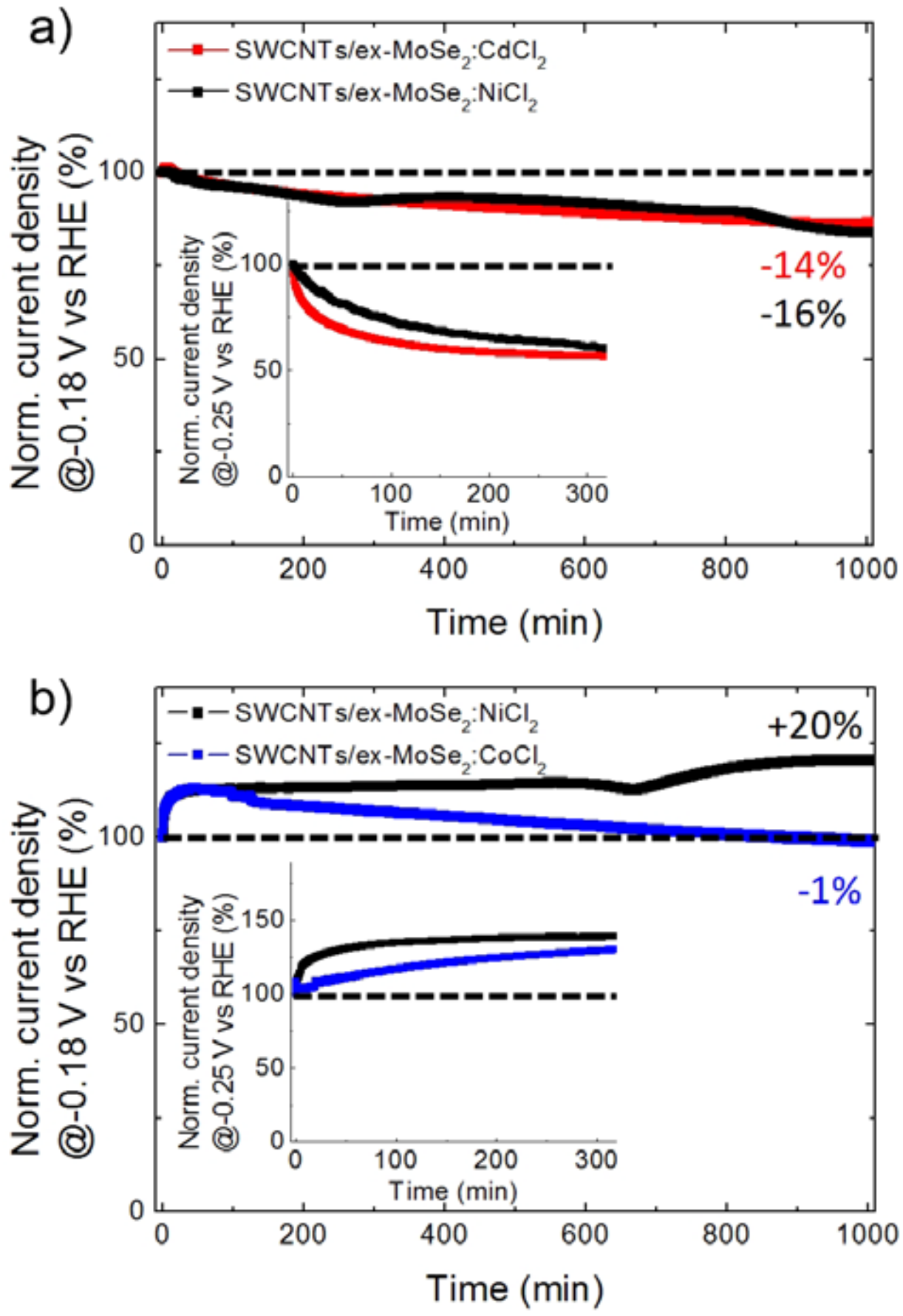

**Figure 5. Chronoamperometry measurements (j-t curves) at a fixed potential of -0.18 V *vs.* RHE for**: a) SWCNTs/ex-$MoSe_2$:$CdCl_2$ and SWCNTs/ex-$MoSe_2$:$NiCl_2$ in 0.5 M $H_2SO_4$; b) SWCNTs/ex-$MoSe_2$:$NiCl_2$ and SWCNTs/ex-$MoSe_2$:$CoCl_2$ in 1 M KOH (f). The insets of panels show the chronoamperometry measurements of ex-$MoSe_2$:$CdCl_2$ and ex-$MoSe_2$:$NiCl_2$ deposited on glassy carbon in 0.5 $H_2SO_4$ (e); ex-$MoSe_2$:$NiCl_2$ and ex-$MoSe_2$:$CoCl_2$ deposited on glassy carbon in 1 M KOH (f).

### 2.3 Hydrogen evolution reaction-activity explanation

In order to explain the HER-assistive role of both the $MCl_2$-doping and the heterostructure design, we hypothesize multiple synergistic functional effects (**Figure 6**).

Concerning the $MCl_2$-doping, these effects include: 1) the modulation of the electronic state of ex-$MoSe_2$, which will improve its intrinsic electrical conductivity (passing from ~10 GΩ cm for ex-$MoSe_2$ film to 700 kΩ cm for the the representative doped film of ex:$MoSe_2$:$NiCl_2$, as estimated by van der Pauw resistivity measurements), *i.e.* the fastening charge transfer towards the HER-active sites (Figure 6a); 2) an increase in the electronic density of the ex-$MoSe_2$ and the local upward band-bending emergence at the p-doped/undoped regions of ex-$MoSe_2$:$MCl_2$ (see XPS analysis, Figure 1e,f), tailoring the capability of the ex-$MoSe_2$ to adsorb H atoms, which will thermodynamically activate the HER (Figure 6b,c);[9,22,24] 3) (in alkaline solution) the promotion of $H_2O$ discharge due to the electrocatalytic activity of metal oxide and hydr(oxy)oxide clusters, formed from $M^0\downarrow$ after material exposure to oxygen/moisture (see XPS analysis, Figure 1g) or under HER-operative electrochemical conditions (which is in agreement with the corresponding Pourbaix diagrams)[122,123] (Figure 6d).[103,124]

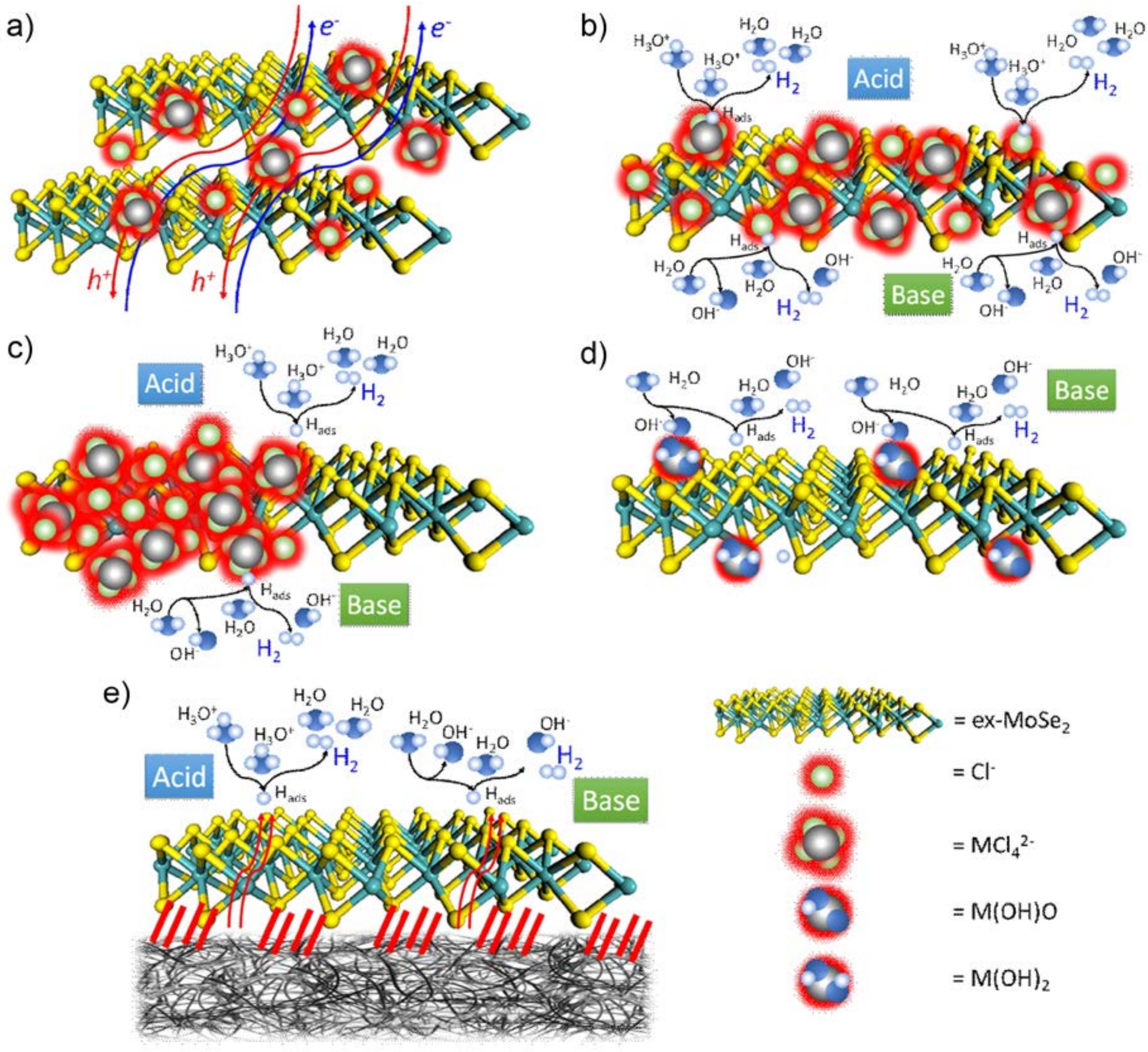

**Figure 6. The HER-assisting effects induced by the $MCl_2$-doping and heterostructure design.** a) The improvement in the electrical conductivity within the ex-$MoSe_2$ film, *i.e.* the fastening charge transfer between the basal planes towards the HER-active sites. b) the creation of new HER-active sites ascribed to the complexes between ex-$MoSe_2$ and $Cl^-$- or the $MCl_4^-$. c) the facilitated production of $H_{ads}$ (in both acidic and alkaline media) and the $H_2O$ discharge (in alkaline media) due the local upward band-bending emergence at the doped/undoped regions of ex-$MoSe_2$. d) the promotion of $H_2O$ discharge in alkaline medium due to the electrocatalytic activity of metal hydr(oxy)oxide clusters formed from $M^0\downarrow$ after exposure to air or under HER-operative electrochemical condition. e) the μm-spatial range HER-assisting electrochemical coupling between SWCNTs and $MoSe_2$ flakes.

Contrary to these beneficial effects, some dopants can also negatively affect the HER-activity of ex-$MoSe_2$. First, in acidic media, the formation of metal oxides and metal hydr(oxy)oxides may passivate the HER active sites for the discharge of $H_3O^+$ and the formation of $H_{ads}$, thus causing a Volmer reaction-limited HER. Secondly, in metal hydr(oxy)oxides, clusters promote $H_2O$ discharge in alkaline media by adsorbing OH on themselves along the $H_{ads}$ formation on the HER-electrocatalysts.[103,124] However, the rate of the desorption of $OH_{ads}$ to regenerate the $H_2O$ dissociative properties is critical in order to determine the resulting HER-activity.[103] In more detail, the decrease in the interaction strength between the $OH_{ads}$ and the M site of the metal hydr(oxy)oxides has been correlated to the HER activity increase in metal hydr(oxy)oxide/Pt (following the order Ni > Co > Fe).[103] Therefore, the HER-activity of ex-$MoSe_2$:$MCl_2$ in alkaline media is expected to have a similar dependence on M to that of metal hydr(oxy)oxide/Pt. These expectations are in agreement with our experimental measurements, where $NiCl_2$ and $FeCl_2$ dopants have led to a maximum and minimum, respectively, ex-$MoSe_2$ HER-activity at pH 14 (Figure 2).

By optimizing the amount of $MCl_2$ doping and investigating different ex-$MoSe_2$:$MCl_2$ ratios during the formulation of the dispersions, the best trade-off between beneficial and detrimental doping-induced effects for HER may be achieved that the HER-activity of the ex-$MoSe_2$:$MCl_2$ flakes can be boosted further.

Concerning the heterostructures' design (*i.e.* the design of SWCTNs/ex-$MoSe_2$:$MCl_2$), the corresponding enhancement of the ex-$MoSe_2$ HER-activity is rationalized by the electrochemical coupling that is established between SWCNTs and $MoSe_2$ (Figure 6e), which causes a decrease in the $\Delta G_H^0$ of the ex-$MoSe_2$ flakes, as has been demonstrated by theoretical studies on carbon-based material/TMD hybrids[89.90]. It is worth noting that the electrochemical coupling in heterostructures is preserved over a longer spatial range (*i.e.* μm-scale) than that of hybrid compounds,[28-31,89,90] as has recently been observed in our previous experimental works.[87,88] In addition, different from the case of carbon-based nanomaterial/TMD hybrids, the heterostructured-based approach for designing HER-electrocatalysts permits each nanomaterial composing the overall electrocatalytic heterostructure to be independently processed. This simplifies the optimization of the material exfoliation protocol and ink formulation. Moreover, it is noteworthy that the use SWCNT paper as a substrate eliminates the implementation of an electrocatalytic film transfer on rigid conductive current collectors (*i.e.* glassy carbon or metal-based substrates),[125] which have to fulfil the electrochemical stability requirements.

## 3. Conclusion

In this work, we have demonstrated that the $MCl_2$-doping of ex-$MoSe_2$ (*i.e.* the production of doped-$MoSe_2$ nanoflakes/3*d* metal oxide-hydr(oxy)oxides hybrids) and the design of heterostructures between SWCNTs and ex-$MoSe_2$:$MCl_2$ are effective techniques with regard to designing pH-universal HER-electrocatalysts. In particular, the use of ex-$MoSe_2$ produced by the cost-effective LPE of the bulk counterpart in an environmentally friendly solvent (*i.e.*, IPA), the $MCl_2$-chemical doping and a solution-processed manufacturing of the heterostructured electrodes, manage to fulfil the key-requirements for pH-universal large-scale $H_2$ production. In fact, the as-produced electrodes reached low $\eta_{10}$ of 0.081 V and 0.064 V in 0.5 M $H_2SO_4$ and 1 M KOH, respectively. The optimization of the heterostructure design (*e.g.* the material mass loading and layer thickness), heterostructure-stacking (*e.g.* the number of heterostructures) and $MCl_2$-doping (*e.g.* the ex-$MoSe_2$:$MCl_2$ molar ratio of the ex-

$MoSe_2$:$MCl_2$ dispersion and the $MCl_2$-doping of SWCNT paper) are expected to further enhance the HER-performance herein achieved. Prospectively, our strategy to hybridize 3*d* metal oxide-hydr(oxy)oxide with HER-active electrocatalysts for enhancing the electrochemical perfomance of the native materials could be effective also for other HER-active low-dimensional materials beyond $MoSe_2$, including metal oxides[126,127,128,129,130]/phospides[131,132]/carbides[133,134]/nitrides[134,135]], as well as other TMDs.[136,137,138] Therefore, we believe that our technology provides an effective low-cost alternative to the current state-of-the-art electrocatalytic production of $H_2$ in both acidic and alkaline media.

## 4. Experimental Section

*Synthesis of materials*

The ex-$MoSe_2$ was produced by LPE,[44,45] followed by SBS,[47-49] in IPA of $MoSe_2$ bulk (crystal, 99.995%, Sigma Aldrich). In short, 30 mg of $MoSe_2$ bulk were added to 50 mL of IPA and ultrasonicated in a bath sonicator (Branson® 5800 cleaner, Branson Ultrasonics) for 6 h. The resulting dispersion was ultracentrifuged at 2700 g (Optima™ XE-90 with a SW32Ti rotor, Beckman Coulter) for 60 min at 15 °C in order to separate un-exfoliated bulk $MoSe_2$ (collected as sediment) from the ex-$MoSe_2$ that remained in the supernatant. Then, the 80% of the supernatant was collected by pipetting, obtaining an ex-$MoSe_2$ dispersion. The concentration of this ex-$MoSe_2$ dispersion was adjusted to 0.4 g $L^{-1}$ by evaporating any excess of the solvent.

The ex-MoSe2:$MCl_2$ dispersions were prepared by dissolving $MCl_2$ ($FeCl_2$: anhydrous, 99.99% trace metals basis, Sigma Aldrich; $CoCl_2$: anhydrous, 99.99%, Sigma Aldrich; $NiCl_2$: anhydrous, 99.99%, Sigma Aldrich; $CuCl_2$: anhydrous, 99.99%, Sigma Aldrich; $ZnCl_2$: anhydrous, 99.99%, Sigma Aldrich; $CdCl_2$: anhydrous, 99.99%, Sigma Aldrich) in the as-obtained ex-$MoSe_2$ dispersion in IPA (ex-$MoSe_2$:$MCl_2$ molar ratio = 1:1). The resulting dispersions were sonicated in a bath sonicator (Branson® 5800 cleaner, Branson Ultrasonics) for 30 min.

The SWCNTs (> 90% purity, Cheap Tubes) were used as-received, without any purification step. The SWCNT dispersions were produced by dispersing SWCNTs in NMP at a concentration of 0.2 g $L^{-1}$ and using ultrasonication-based de-bundling[139,140,141]. Briefly, 10 mg of SWNT powder was added to 50 mL of IPA in a 100 mL open top, flat-bottomed beaker. The dispersion was sonicated using a horn probe sonic tip (Vibra-cell 75185, Sonics) with a vibration amplitude set to 45% and a sonication time of 30 min. The sonic tip was pulsed for 5 s on and 2 s off to avoid damage to the processor and to reduce any solvent heating. An ice bath around the beaker was used during sonication in order to minimize any heating effects.

The Pt/C powder (20 wt% platinum on carbon black, Alfa Aeasar) was used as-purchased. The Pt/C dispersion were prepared by dispersing Pt/C powder in 1 mL water/ethanol mixed solvent (v/v = 1 : 1) at a concentration of 5 mg $mL^{-1}$. Then, 10 μL of Nafion® 117 solution (~5% in a mixture of lower aliphatic alcohols and water, Sigma Aldrich) were added into the Pt/C dispersion. The so-obtained dispersion was sonicated for 30 min before the use.

*Characterization of materials*

Transmission electron microscopy images were taken with a JEM 1011 (JEOL) transmission electron microscope, operating at 100 kV. Morphological and statistical analysis was carried out by using ImageJ software (NIH) and OriginPro 9.1 software (OriginLab), respectively. Samples for the TEM measurements were prepared by drop-casting the ex-$MoSe_2$ dispersion onto carbon-coated Cu grids, rinsing them with deionized water and subsequently drying them under vacuum overnight.

Atomic force microscopy images were taken using a Nanowizard III (JPK Instruments, Germany) mounted on an Axio Observer D1 (Carl Zeiss, Germany) inverted optical microscope. The AFM measurements were carried out by using PPP-NCHR cantilevers (Nanosensors, USA) with a nominal tip diameter of 10 nm. A drive frequency of ~295 kHz was used. Intermittent contact mode AFM images (512×512 data points) of 2.5×2.5 μm$^2$ and 500×500 nm$^2$ were collected by keeping the working set point above 70% of the free oscillation amplitude. The scan rate for the acquisition of images was 0.7 Hz. Height profiles were processed by using the JPK Data Processing software (JPK Instruments, Germany) and the data were analysed with OriginPro 9.1 software. Statistical analysis was carried out by means of Origin 9.1 software on four different AFM images for each sample. The samples were prepared by drop-casting ex-$MoSe_2$ dispersions onto mica sheets (G250-1, Agar Scientific Ltd., Essex, U.K.) and drying them under vacuum.

X-ray photoelectron spectroscopy characterization was carried out on a Kratos Axis UltraDLD spectrometer, using a monochromatic Al Kα source (15 kV, 20 mA). The spectra were taken on a 300×700 μm$^2$ area. Wide scans were collected with a constant pass energy of 160 eV and an energy step of 1 eV. High-resolution spectra were acquired at constant pass energy of 10 eV and an energy step of 0.1 eV. The binding energy scale was referenced to the C 1*s* peak at 284.8 eV. The spectra were analyzed using CasaXPS software (version 2.3.17). The fitting of the spectra was performed by using a linear background and Voigt profiles. The samples were prepared by drop-casting ex-$MoSe_2$ and ex-$MoSe_2$:$MCl_2$ dispersions onto a $Si/SiO_2$ substrate (LDB Technologies Ltd) and drying them under vacuum.

Raman spectroscopy measurements were carried out using a Renishaw microRaman invia 1000 with a 50× objective, an excitation wavelength of 532 nm and an incident power on the samples of 1 mW. For each sample, 50 spectra were collected. The samples were prepared by drop casting ex-$MoSe_2$ and ex-$MoSe_2$:$MCl_2$ dispersions on $Si/SiO_2$ substrates and drying them under vacuum. The spectra were fitted with Lorentzian functions. Statistical analysis was carried out by means of OriginPro 9.1 software.

Van der Pauw resistivity measurement were acquired on ex-$MoSe_2$ and ex-$MoSe_2$:$NiCl_2$ films with an average thickness of ~1.1 μm (mass loading of 0.8 mg cm$^{-2}$). The material films were obtained by dropcasting the materials dispersions onto glass substrates. Metal contacts in van der Pauw geometry were made with Ag paste, and current-voltage curves were measured with a Keithley 2600 SMU.

*Fabrication of electrodes*

The ex-$MoSe_2$ and ex-MoSe2:$MCl_2$ were deposited on GC sheets (Sigma Aldrich) (electrodes labelled ex-$MoSe_2$ and ex-$MoSe_2$:$MCl_2$) by drop-casting the as-produced ex dispersion (mass loading of 0.2 mg $cm^{-2}$ referred to ex-$MoSe_2$). The SWCNTs/ex-$MoSe_2$ (or SWCTNs/ex-$MoSe_2$:$MCl_2$) were fabricated by depositing the SWCTN and ex-$MoSe_2$ (or ex-$MoSe_2$:$MCl_2$) dispersion onto nylon membranes (Whatman® membrane filters nylon, 0.2 μm pore size, Sigma Aldrich) *via* a vacuum filtration process.[87,88,142] The mass loading of the materials was 0.64 mg $cm^{-2}$ for both SWCNTs and ex-$MoSe_2$ (for ex-$MoSe_2$:$MCl_2$ this value refers to the ex-$MoSe_2$). The electrode area was 3.14 $cm^2$. The Pt/C electrodes were realized by depositing Pt/C dispersion on GC sheets (mass loading of 0.35 mg $cm^{-2}$). All the electrodes were dried overnight at room temperature before their electrochemical characterization.

*Characterization of electrodes*

Scanning electron microscope analysis was performed with a field-emission scanning electron microscope FE-SEM (Jeol JSM-7500 FA). The acceleration voltage was set to 5kV. Images were collected using a secondary electron sensor for LEI images and the in-lens sensor for SEI images. Energy-dispersive X-ray spectroscopy images were acquired at 5kV by a silicon drift detector (Oxford Instruments X-max 80) having an 80 $mm^2$ window. The EDX analysis was performed using Oxford Instrument AZtec 3.1 software. The SWCNTs/ex-$MoSe_2$:$MCl_2$ with M = Ni, Co, and Cd were analysed as representative cases.

Electrochemical measurements were carried out at room temperature in a flat-bottom fused silica cell under a three-electrode configuration using CompactStat potentiostat/galvanostat station (Ivium), controlled via Ivium's own IviumSoft. A Pt wire was used as the counter-electrode and saturated KCl Ag/AgCl was used as the reference electrode. Measurements were carried out in 200 mL of 0.5 M $H_2SO_4$ (99.999% purity, Sigma Aldrich) or 1 M KOH (≥ 85% purity, ACS reagent, pellets, Sigma Aldrich). Oxygen was purged from the electrolyte by flowing $N_2$ gas through the liquid volume using a porous frit for 30 min before starting the measurements. A constant $N_2$ flow was maintained afterwards for the whole duration of the experiments, to avoid the re-dissolution of molecular oxygen in the electrolyte. The potential difference between the working electrode and the Ag/AgCl reference electrode was converted to the reversible hydrogen electrode (RHE) scale *via* the Nernst equation: $E_{RHE} = E_{Ag/AgCl} + 0.059 \times pH + E^0_{Ag/AgCl}$, in which $E_{RHE}$ is the converted potential *vs.* RHE, $E_{Ag/AgCl}$ is the experimental potential measured against the Ag/AgCl reference electrode, and $E^0_{Ag/AgCl}$ is the standard potential of Ag/AgCl at 25 °C (0.1976 V *vs.* RHE). The LSV curves were acquired at the scan rate of 5 mV $s^{-1}$. Polarization curves from all catalysts were *iR*-corrected, in which *i* is the measured working electrode current and the *R* is the series resistance that arises from the working electrode substrate and electrolyte resistances. *R* was measured by electrochemical impedance spectroscopy (EIS) at open circuit potential and at the frequency of $10^4$ Hz.

The linear portions of the Tafel plots fitted the Tafel equation $\eta = b*|\log(j)| + A$,[85,143] in which η is the overpotential in comparison with the reversible hydrogen electrode potential (0 V *vs.* RHE), j is the current density, *b* is the Tafel slope and *A* is the intercept of the linear regression. By setting η equal to zero in the Tafel equation, the $j_0$ was calculated. Stability tests were carried out by chronoamperometry measurements (j-t curves), *i.e.* by measuring the current in the potentiostatic mode at a fixed overpotential of 0.25 V for ex-$MoSe_2$:$MCl_2$ (GC substrate) and 0.18 V for SWCNT/ex-$MoSe_2$:$MCl_2$ (heterostructure) in 0.5 M $H_2SO_4$ over time.

Elemental analysis by inductively coupled plasma optical emission spectroscopy was performed with a ThermoFisher® ICAP 6000 Duo inductively coupled plasma optical emission spectrometer. The samples were digested solutions prepared from the 0.5 M $H_2SO_4$ solutions (volume of 100 mL) recovered after carrying out the HER measurements of SWCTNs/ex-$MoSe_2$:$CdCl_2$ or SWCNTs/ex-$MoSe_2$:$NiCl_2$. To avoid incompatibilities of the ICP-OES system with the presence of $H_2SO_4$, the aliquot (3 mL) used for the analysis was neutralized to pH = 7 using $NH_4OH$ (5 M, Fluka®, 120 uL). Following this procedure, no metals were added in the solution. This was confirmed by the blank prepared just using water and the same amount of $NH_4OH$ solution. Then, two series of samples together with the corresponding blank were prepared in a 25 mL volumetric flask, digesting the neutralized aliquot in 2 mL of aqua regia (HCl:$HNO_3$, volume ratio of 3:1) overnight and 2 days, respectively. Prior to the measurement, each sample was diluted by 25 mL with Millipore water to the total volume, and stirred by vortex at 2400 rpm for 10 s. Lastly, the sample was filtered using a PTFE membrane (0.45 μm pore size). Three measurements were performed on each sample to obtain the final averaged values of metal concentration.

**Acknowledgements**

This project has received funding from the European Union's Horizon 2020 research and innovation program under grant agreement No. 696656—GrapheneCore1. We thank: Beatriz Martín-García for her support in ICP-MS measurements; Lea Pasquale (Materials Characterization Facility, Istituto Italiano di Tecnologia) for her support in XPS data acquisition; Electron Microscopy facility, Istituto Italiano di Tecnologia for support in SEM/TEM data acquisition; Davide Spirito for his support in van der Pauw resistivity measurements; Nanochemistry facility, Istituto Italiano di Tecnologia for support facility for support in electrochemical measurements and electrode preparation.
L. Najafi and S. Bellani contributed equally to this work.

# Supporting Information

**Doped-$MoSe_2$ nanoflakes/3*d* metal oxide-hydr(oxy)oxides hybrid catalysts for pH-universal electrochemical hydrogen evolution reaction**

*Leyla Najafi, Sebastiano Bellani, Reinier Oropesa-Nuñez, Alberto Ansaldo, Mirko Prato, Antonio Esau Del Rio Castillo and Francesco Bonaccorso**

## S1. X-ray photoelectron spectroscopy analysis of the oxidation level of Mo for ex-$MoSe_2$:$MCl_2$

The X-ray photoelectron spectroscopy (XPS) measurements for the ex-$MoSe_2$ and ex-$MoSe_2$:$MCl_2$ are reported in Figure 1e-g of the main text. The Mo 3*d* spectra (Figure 1e) show two peaks located at ~229 eV and ~232 eV which are attributed to the Mo $3d_{5/2}$ and Mo $3d_{3/2}$ peaks of the $Mo^{4+}$ state in $MoSe_2$.[1,2] The peaks at higher binding energies (*i.e.* ~232 eV and ~236 eV) are assigned to the $Mo^{6+}$ state, and are attributed to the $MoO_3$ residues in pristine materials. The chemical $MCl_2$-doping of the ex:$MoSe_2$ and the formation of interfacial dipole complexes between doped-ex-$MoSe_2$ and metal oxides/hydr(oxy)oxides alter the electronic density of the ex-$MoSe_2$ (as discussed in the main text),[3] causing a different level of oxidation for Mo in ex-$MoSe_2$. In particular, chemical composition analysis evidences that the percentage content (%c) of $Mo^{6+}$ increased from 23% in ex-$MoSe_2$ to 75%, 26%, 42%, 25%, 47%, and 58% after doping with $FeCl_2$, $CoCl_2$, $NiCl_2$, $CuCl_2$, $ZnCl_2$ and $CdCl_2$, respectively. Notably, $FeCl_2$-doping led to the highest level of oxidation of ex-$MoSe_2$. Furthermore, only in this case, the ratio between the %c of Cl and that of M is lower than 1. This is tentatively explained by taking into account the evolution of HCl, which have been formed when residual $FeCl_2$ reacts with $O_2$ or $H_2O$ to form metal hydroxyoxide (*i.e.* $4MCl_2 + 6H_2O + O_2 \rightleftarrows 4 M(OH)O + 8HCl$)[4,5,6] or metal hydroxychloride (*i.e.* $MCl_2 + H_2O \rightleftarrows M(OH)Cl + HCl$)[4-6], respectively, during the drying in air of ex-$MoSe_2$:$FeCl_2$, which had previously been deposited onto a Si/$SiO_2$ substrate.

A binding energy peak of Cl 2*p* at ~198 eV is also detected in the ex-$MoSe_2$:$MCl_2$ (data not shown), together with distinct binding energy peaks that are associated with the M 2*p* or 3*d* doublets (Figure 1g) (Fe $2p_{3/2}$ at 711 eV (satellite features at 720 eV) and Fe $2p_{1/2}$ at 725 eV; Co $2p_{3/2}$ at 882 eV (satellite feature at 887 eV) and Co $2p_{1/2}$ at 798 eV (satellite feature at 803 eV); Ni $2p_{3/2}$ at 856 eV (satellite feature at 863 eV) and Ni $2p_{1/2}$ at 874 eV (satellite feature at 881 eV); Ni $2p_{3/2}$ at 856 eV (satellite feature at 863 eV) and Ni $2p_{1/2}$ at 874 eV (satellite feature at 881 eV); Cu $2p_{3/2}$ at 932 and 935 eV (satellite features between 942 and 946 eV) and Cu $2p_{1/2}$ at 952 and 955 eV (satellite features around 963 eV); Zn $2p_{3/2}$ at 1023 eV and Zn $2p_{1/2}$ at 1046 eV; Cd $3d_{5/2}$ at 406 eV and Cd $3d_{3/2}$ at 413 eV. All the peaks refer to the oxidized states of M (*i.e.* $Fe^{3+}$, $Co^{2+}$, $Ni^{2+}$, $Cu^{1+}$ or $Cu^{2+}$, $Zn^{2+}$ and $Cd^{2+}$), which are attributed to both $MCl_2$ residuals and oxidized species formed by the reaction between $M^0\downarrow$ and $O_2$ or $H_2O$, after the exposure to ambient. The oxidized species can be either metal oxide (*e.g.* MO for $M^{2+}$, $M_2O_3$ for $M^{3+}$) or metal hydr(oxy)oxides (*e.g.* $M(OH)_2$ for $M^{2+}$ or $M(OH)_3$ and M(OH)O for $M^{3+}$. For the case of Cu, the additional peaks are also observed at 932 and 952 eV, as ascribed to the Cu $2p_{3/2}$ and Cu $2p_{1/2}$ of $Cu^{1+}$ species (*e.g.* $Cu_2O$).

**S2. Raman Statistical analysis of $MoSe_2$ bulk and ex-$MoSe_2$**

**Figure S1** shows the statistical Raman analysis of Pos ($A_{1g}$) of $MoSe_2$ bulk (panel a) and Pos($A_{1g}$), Pos($E^1_{2g}$) I($A_{1g}$)/I($E_{2g}$) of ex-$MoSe_2$ (panel b, c and d, respectively). As discussed in the main text, the $A_{1g}$ mode is located at ~241 $cm^{-1}$ for the $MoSe_2$ bulk, while it red-shifts to ~239 $cm^{-1}$ for the ex-$MoSe_2$, which is in agreement with the softening of the vibrational mode accompanied by the reduction in layer thickness.[7,8,9,10,11] The intensity ratio between the $A_{1g}$ and $E^1_{2g}$ modes (I($A_{1g}$)/I($E^1_{2g}$)) for ex-$MoSe_2$ is ∼21. This value agrees with those reported for few-layer $MoSe_2$ flakes.[12]

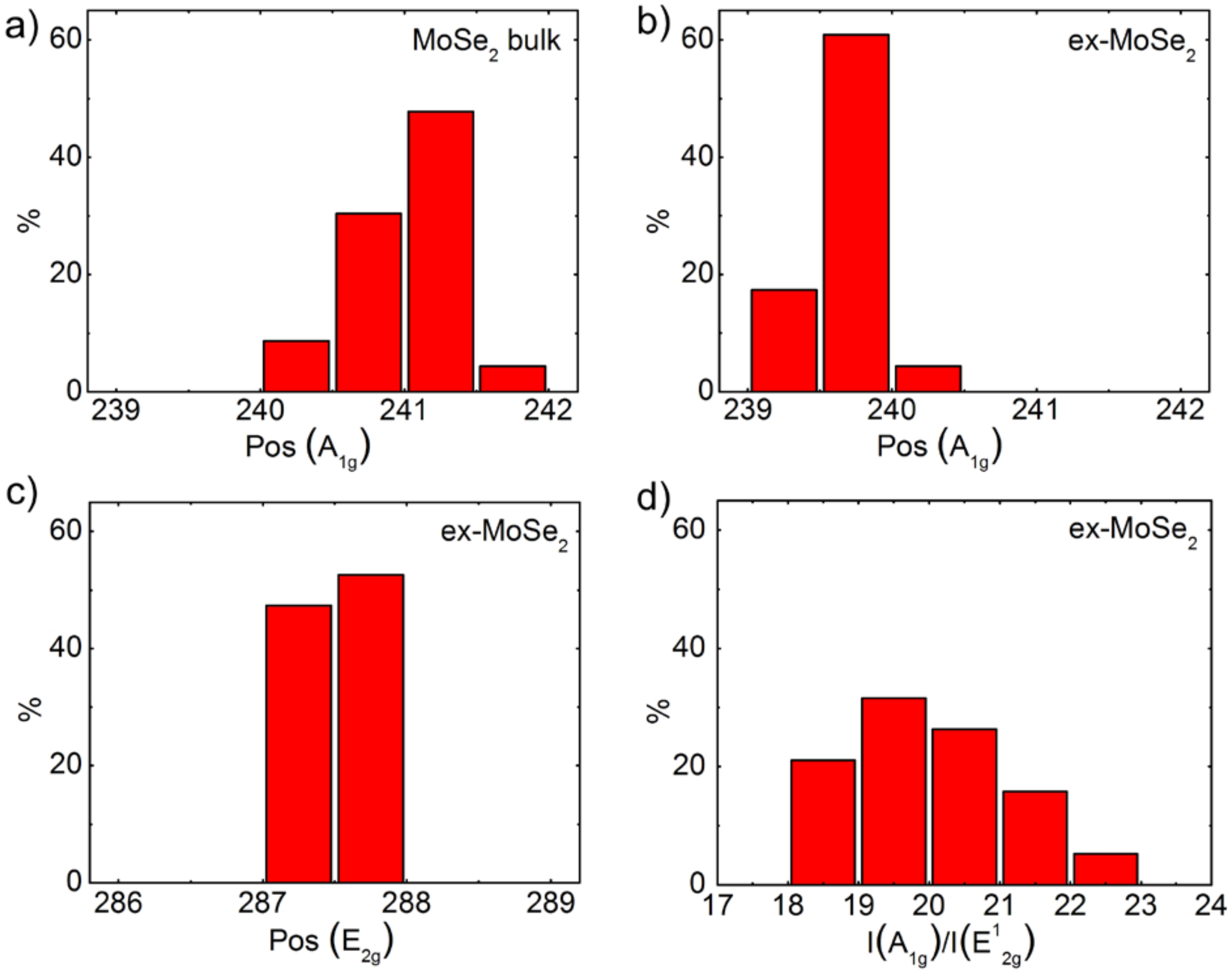

**Figure S1. Raman statistical analysis of $MoSe_2$ bulk and ex-$MoSe_2$.** a,b) Statistical Raman analysis of Pos ($A_{1g}$) of $MoSe_2$ bulk and ex-$MoSe_2$. c,d) Statistical Raman analysis of Pos($E_{2g}$) and I($A_{1g}$)/I($E_{2g}$) of ex-$MoSe_2$.

**S3. Exchange current densities of the ex-$MoSe_2$ and ex-$MoSe_2$:$MCl_2$ in acidic and alkaline solutions**

**Figure S2** shows the exchange current densities ($j_0$) in both acidic (0.5 M $H_2SO_4$) and alkaline (1 M KOH) solutions, as extrapolated from the analysis of the Tafel plots (Figure 2c,d in the main text). Under acidic conditions, the $j_0$ was significant affected by $NiCl_2$-doping ($j_0$ ~292 μA $cm^{-2}$), but there were no remarkable variation for the others $MCl_2$-doping ($j_0$ in the 10-100 μA $cm^{-2}$ range). Under alkaline conditions, the ex-$MoSe_2$ and the ex-$MoSe_2$:$MCl_2$ showed comparable $j_0$ values (in the range of 10-30 μA $cm^{-2}$) (with the only exception being ex-$MoSe_2$:$CuCl_2$ which displayed a higher $j_0$ value of ~70 μA $cm^{-2}$).

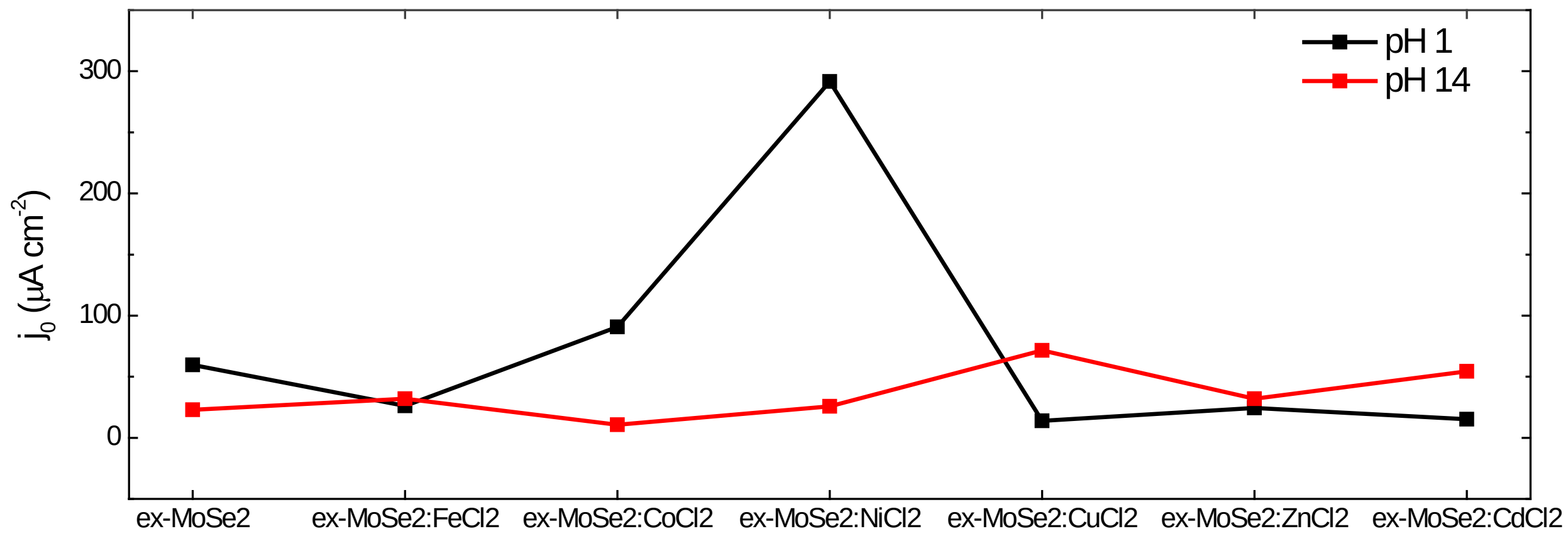

**Figure S2. Exchange current densities ($j_0$) analysis of the ex-$MoSe_2$ and ex-$MoSe_2$:$MCl_2$.** The $j_0$ values under acidic (black squares) and alkaline conditions (red squares) for the different materials, as calculated for the analysis of the Tafel plots reported in the Figure 2c,d of the main text.

**S4. AFM analysis of the SWCNT paper**

**Figure S3** reports the AFM images of the SWCNTs deposited onto nylon membranes from their as-produced dispersions *via* a vacuum filtration process (SWCNT mass loading: 0.64 mg $cm^{-2}$). The surface of the SWCNT paper resembles a network with a bundle-like arrangement, which is consistent with the SEM analysis shown in the main text (Figure 3a). The estimated roughness average (Ra) value is ~103 nm. As explained in the main text, this value decreases to 70 nm after ex-$MoSe_2$ deposition, indicating the ex-$MoSe_2$ overlay leads to a surface flattening.

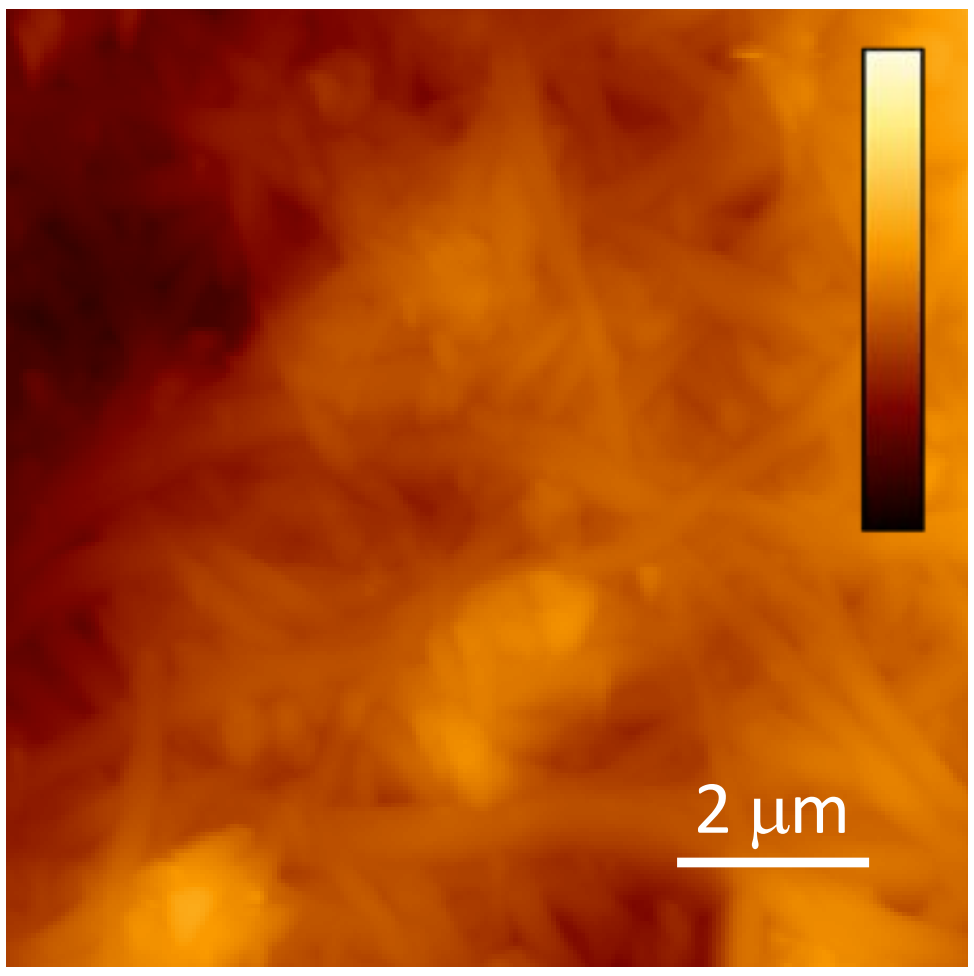

**Figure S3. Topography characterization of the SWCNT paper.** AFM image of the surface of the SWCNT paper. The estimated Ra values is ~103 nm. The z-scale bar is 1 µm.

## S5. Elemental energy-dispersive X-ray spectroscopy (EDX) analysis of cross-sectional SEM images of SWCNTs/ex-MoSe2:$MCl_2$

**Figure S4**a shows the EDX analysis of a top-view SEM image of SWCNTs/ex-$MoSe_2$:$NiCl_2$ (atom color code: yellow C; cyan Mo; violet Se, red Ni), while Figure S4b reports the corresponding mass spectrum. Figure S4c-f report the corresponding EDX analysis for the C, Mo, Se and Ni atoms, respectively. These results evidence that Ni was uniformly distributed over the heterostructure surface, thus indicating that the $MCl_2$-doping of ex-$MoSe_2$ was uniform.

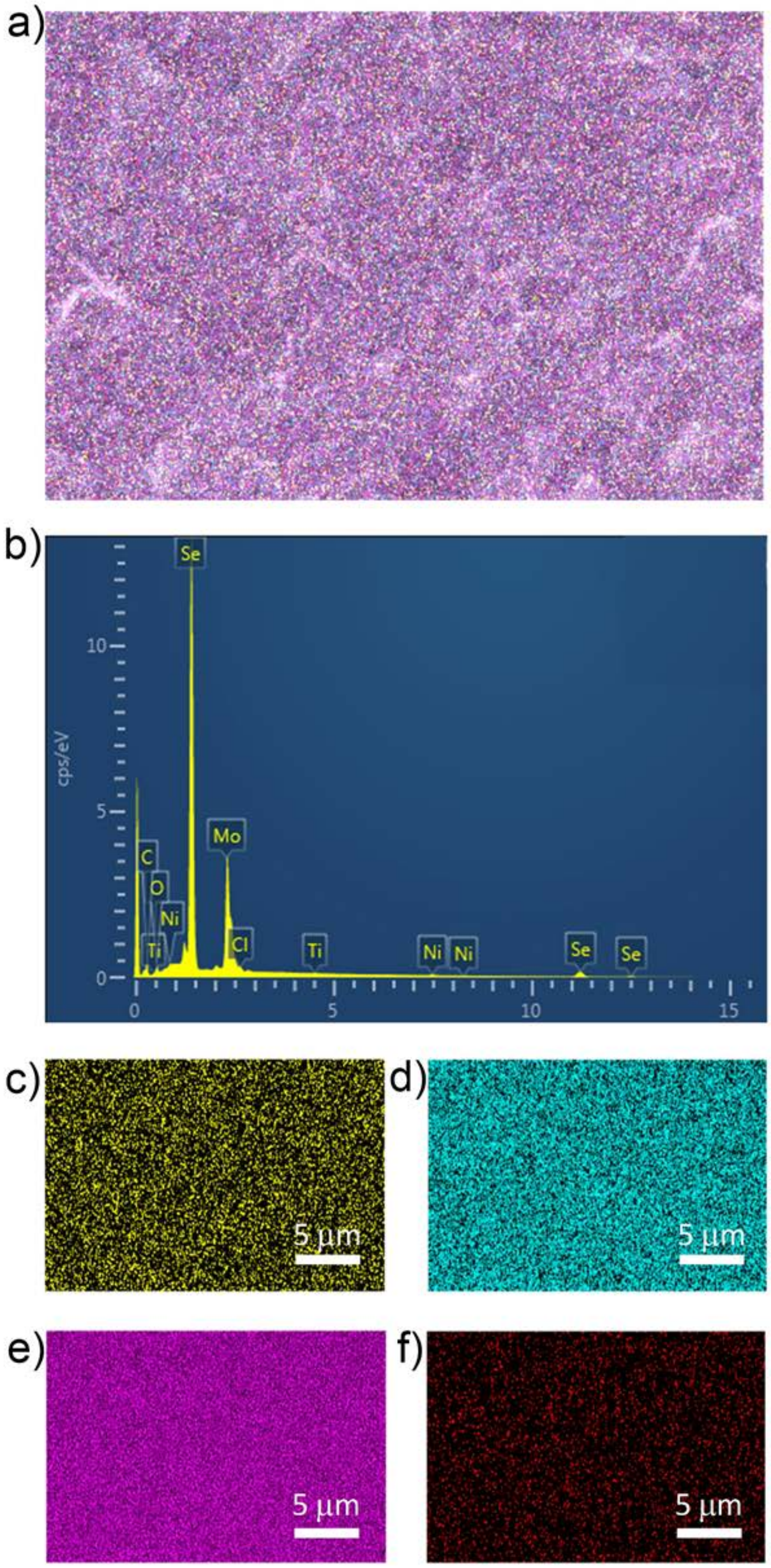

**Figure S4. Elemental energy-dispersive X-ray spectroscopy (EDX) analysis of the surface of SWCNTs/ex-MoSe$_2$:NiCl$_2$.** a) Elemental mapping of a representative top-view SEM image of SWCNTs/ex-MoSe$_2$:NiCl$_2$. b) The mass spectrum of the EDX analysis of the SEM image in panel (a). c-f) EDX analysis of the SEM image showed in panel a for a single element: c) C, d) Mo, e), Se and f) Ni. Atom color code: yellow C; cyan Mo; violet Se; red Ni.

**Figure S5**a shows the EDX analysis, of a representative cross-sectional SEM image of SWCNTs/ex-$MoSe_2$:$NiCl_2$ (same atom color code of Figure S3), while Figure S5b reports the corresponding mass spectrum. Figure S5c-f reports the corresponding EDX analysis for C, Mo, Se and Ni atoms, respectively. These results evidence that, although it have a defined bilayer-like structure, the ex-$MoSe_2$-based overlay partially penetrates into the SWCNT paper. Moreover, Ni is uniformly distributed along the vertical direction of the heterostructure. This means that, during deposition of the ex-$MoSe_2$:$MCl_2$, the formed $M^0$↓ and, eventually, $MCl_2$ residuals can infiltrate the SWCNT paper. Thus, the formation of metal oxide (MO or $M_2O_3$) or metal hydr(oxy)oxides ($M(OH)_2$, $M(OH)_3$ or M(OH)O) and, eventually, the $MCl_2$-doping of the SWCNT paper[13] can occur during the fabrication of the heterostructures. It is worth noting that a p-doping of the SWCNT is beneficial for enhancing the electron transfer from the SWCTN paper to the active sites of the ex-$MoSe_2$:$MCl_2$.[14]

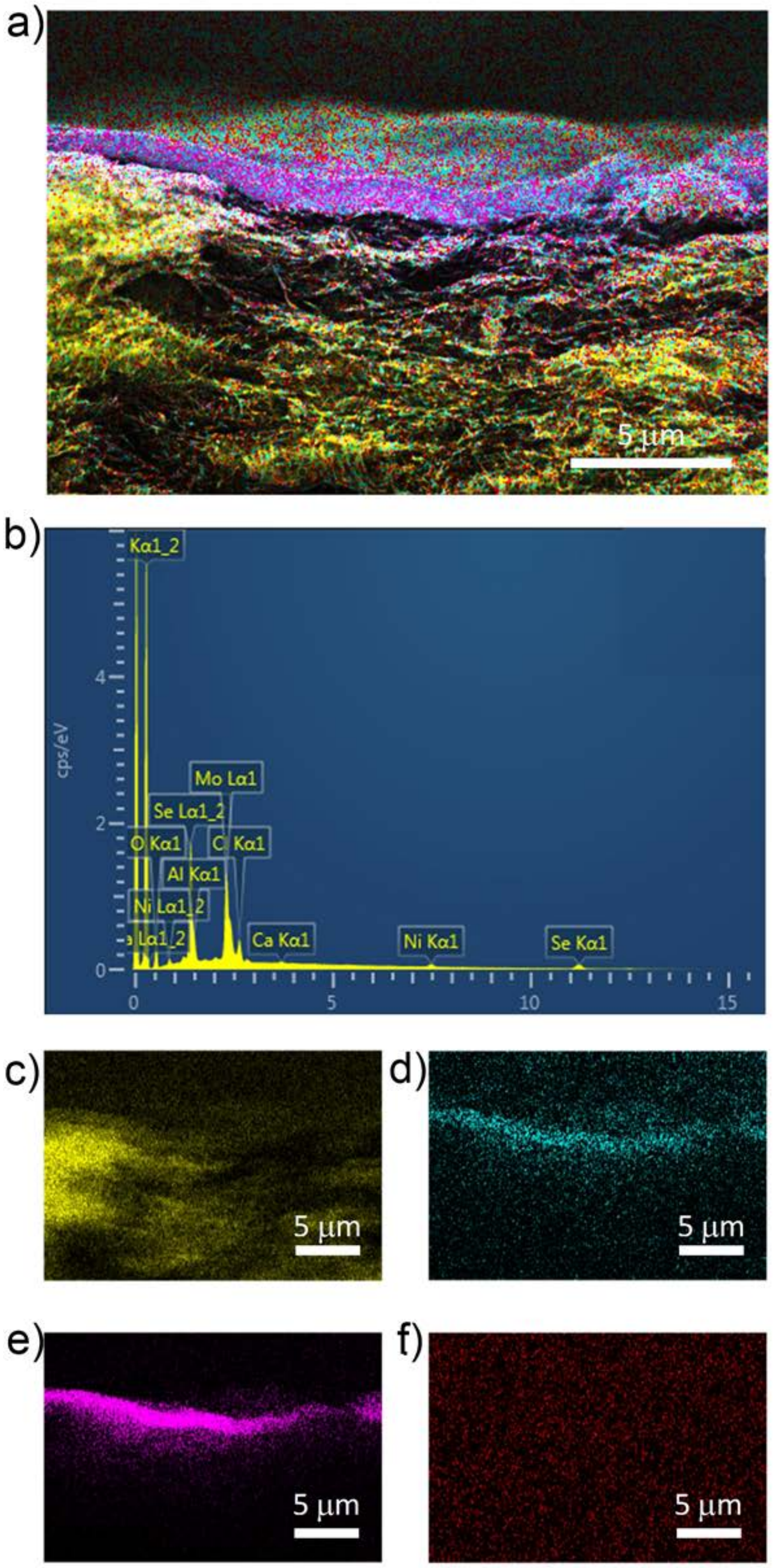

**Figure S5. Elemental energy-dispersive X-ray spectroscopy (EDX) analysis of the cross-section of SWCNTs/ex-$MoSe_2$:$NiCl_2$.** a) Elemental mapping of a representative cross-sectional SEM image of SWCNTs/ex-$MoSe_2$:$NiCl_2$. b) The mass spectrum of the EDX analysis of the SEM image of panel a. c-f) EDX analysis of the SEM image showed in panel a for the single element: c) C, d) Mo, e), Se and f) Ni. Atom color code: yellow C; cyan Mo; violet Se; red Ni.

## S6. X-ray photoelectron spectroscopy of ex-$MoSe_2$ after HER-operation in alkaline condition

**Figure S6** shows the Mo 3*d* and Se 3*d* XPS spectra ex-$MoSe_2$ deposited on glassy carbon after HER measurements in 1 M KOH for 1000 min. Clearly, the data do not evidence any changes in the XPS spectra, thus indicating that no degradation occurred on ex-$MoSe_2$, similarly to that observed in acidic condition.[15]

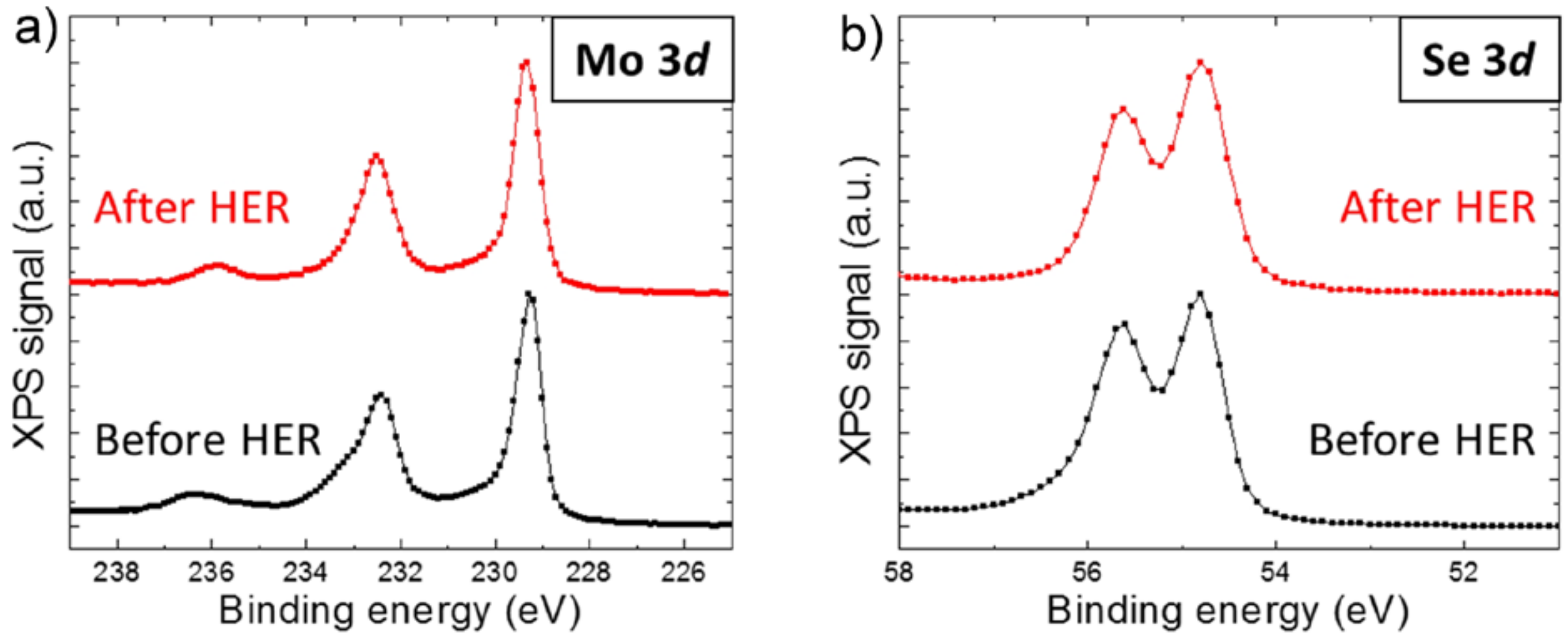

**Figure S6. X-ray photoelectron spectroscopy analysis of ex-$MoSe_2$ before and after HER measurements in alkaline condition.** a) Mo 3*d* and b) Se 3*d* XPS spectra for $MoSe_2$ deposited on glassy carbon before (black dotted lines) and after HER measurements (red dotted lines) in 1 M KOH.

**Table S1** Summary of the HER-activities of the designed electrocatalyst and other relevant noble metal-free electrocatalysts reported in literature.

| Electrocatalyst | Overpotential at 10 mA $cm^{-2}$ ($\eta_{10}$) (V) | Electrolyte |
|---|---|---|
| **4-SWCNTs/$MoSe_2$:$CdCl_2$ (this work)** | **0.081** | **0.5 M $H_2SO_4$** |
| **4-SWCNTs/$MoSe_2$:$NiCl_2$ (this work)** | **0.064** | **1 M KOH** |
| Amorphous $MoS_x$ film[16] | 0.200 at 14 mA $cm^{-2}$ | 1 M $H_2SO_4$ |
| | 0.540 at 4 mA $cm^{-2}$ | 0.1 M KOH |
| Amorphous Ni-$MoS_x$ film[17] | 0.215 at 1 mA $cm^{-2}$ | Phosphate buffer (pH 7) |
| Chemical vapor deposition (CVD)-$MoS_2$ monolayer film[18] | 0.190 at 20 mA $cm^{-2}$ | 0.5 M $H_2SO_4$ |
| $CoS_x$/$MoS_2$ chalcogels[19] | 0.210 at 5 mA $cm^{-2}$ | 0.1 M KOH |
| | 0.235 at 5 mA $cm^{-2}$ | 0.1 M $HClO_4$ |
| Double-gyroid mesoporous $MoS_2$ film[20] | 0.235 | 0.5 M $H_2SO_4$ |
| Metallic $MoS_2$ nanoflakes[21] | 0.187 | 0.5 M $H_2SO_4$ |
| Defect-rich $MoS_2$ nanoflakes[22] | 0.19 | 0.5 M $H_2SO_4$ |
| Li-$MoS_2$ nanoflakes film[23] | 0.168 | 0.5 M $H_2SO_4$ |
| Strained $MoS_2$ nanoflakes[24] | 0.170 | 0.5 M $H_2SO_4$ |
| $MoS_2$ nanoflakes on N-doped carbon nanotube (CNT) forest[[25] | 0.110 | 0.5 M $H_2SO_4$ |
| Edge-terminated $MoS_2$ nanoflakes[26] | 0.149 | 0.5 M $H_2SO_4$ |
| Vertical arrays of $MoS_2$ sheets terminated with such a stepped surface structure[27] | 0.104 | 0.5 M $H_2SO_4$ |
| $MoS_2$ nanoparticles/graphene[28] | 0.155 | 0.5 M $H_2SO_4$ |
| Dimeric $[Mo_3S_{13}]^{2-}$ clusters[29] | 0.180 | 0.5 M $H_2SO_4$ |
| $MoS_2$ nanosheets on hollow carbon spheres[30] | 0.126 | 0.5 M $H_2SO_4$ |
| $MoS_2$ nanoflakes/mesoporous graphene[31] | 0.140 | 0.5 M $H_2SO_4$ |
| Metallic phase $MoS_2$ nanoflakes[32] | 0.175 | 0.5 M $H_2SO_4$ |
| High-Content metallic 1T phase $MoS_2$[33] | 0.126 | 0.5 M $H_2SO_4$ |
| 1T-$MoS_2$/SWNT[34] | 0.108 | 0.5 M $H_2SO_4$ |
| Thermally texturized $MoS_2$ microflakes[35] | 0.174 | 0.5 M $H_2SO_4$ |
| N-doped $MoS_2$[36] | 0.168 | 0.5 M $H_2SO_4$ |
| Ni-doped $MoS_2$[37] | 0.098 | 1 M KOH |
| Zn-doped $MoS_2$[38] | 0.130 | 0.5 M $H_2SO_4$ |
| P-doped $MoS_2$ nanoflakes[39] | 0.043 | 0.5 M $H_2SO_4$ |
| Defect-rich $MoS_2$ nanowalls[40] | 0.095 | 0.5 M $H_2SO_4$ |
| Electrodeposited $MoS_x$ films assisted by liquid crystal template[41] | 0.093 | 0.5 M $H_2SO_4$ |
| $MoS_2$ quantum dots/graphene heterostructure[42] | 0.136 | 0.5 M $H_2SO_4$ |
| $Mo_3S_{13}$ Films[43] | 0.200 | 0.5 M $H_2SO_4$ |
| $MoSe_2$ nanoflakes film[44] | ~0.290 | 0.5 M $H_2SO_4$ |
| $MoSe_2$ nanofilm grown on carbon cloth[45] | 0.220 | 0.5 M $H_2SO_4$ |
| $MoSe_2$ grown on $SnO_2$ nanotubes[46] | 0.174 | 0.5 M $H_2SO_4$ |
| S-doped $MoSe_2$ nanoflakes[10] | ~0.10 | 0.5 M $H_2SO_4$ |
| $MoSe_2$ nanofilm on carbon fiber paper[47] | ~0.25 | 0.5 M $H_2SO_4$ |
| $MoSe_{2-x}$ (x=0.47) nanoflakes[48] | ~0.28 | 0.5 M $H_2SO_4$ |
| Vertical nitrogen-doped 1T-2H $MoSe_2$/graphene shell/core nanoflake arrays[49] | 0.098 | 0.5 M $H_2SO_4$ |
| $MoSe_2$ nanoflakes/reduced graphene oxide (RGO) hybrid[44] | ~0.15 | 0.5 M $H_2SO_4$ |
| Porous $MoSe_2$ nanoflakes[50] | ~0.15 | 0.5 M $H_2SO_4$ |
| $MoSe_2$ nanoflakes perpendicularly grown on CNTs[51] | 0.178 | 0.5 M $H_2SO_4$ |
| $MoSe_2$ nanoflake/CNT heterostructure[15] | 0.100 | 0.5 M $H_2SO_4$ |
| $MoSe_2$ nanocrystal[52] | 0.107 | 0.5 M $H_2SO_4$ |
| $MoSe_2$ nanosheets/N-doped CNT hybrid[53] | 0.102 | 0.5 M $H_2SO_4$ |
| $MoS_{2(1-x)}Se_{2x}$ particles/$NiSe_2$ foam hybrid[54] | 0.069 | 0.5 M $H_2SO_4$ |
| 3D $MoSe_2$@$Ni_{0.85}Se$ nanowire network[55] | 0.117 | 1 M KOH |
| $WS_{2(1-x)}P_{2x}$ nanoribbons[56] | 0.098 | 0.5 M $H_2SO_4$ |
| $WO_3{\cdot}2H_2O$ nanoplates/$WS_2$ hybrid[57] | 0.152 at 100 mA $cm^{-2}$ | 0.5 M $H_2SO_4$ |
| $WS_2$ nanosheets[58] | ~100 | 0.5 M $H_2SO_4$ |
| $Co_xS_y$/$WS_2$ nanosheets on carbon cloth[59] | 0.120 at mA $cm^{-2}$ | 0.5 M $H_2SO_4$ |
| Strained Li-$WS_2$ nanosheets[60] | 80-100 (1T-phase)<br>~200 (2H phase) | 0.5 M $H_2SO_4$ |
| $W_xC$@$WS_2$ nanostructure[61] | 0.146 | 0.5 M $H_2SO_4$ |
| N-doped $Mo_2C$[62] | 0.110 | 1 M KOH |
| $CoSe_2$ nanocrystals embedded into carbon nanowires[63] | 0.130 | 0.5 M $H_2SO_4$ |
| $CoSe_2$ nanoparticles grown on carbon fiber paper [64] | 0.139 | 0.5 M $H_2SO_4$ |
| Porous $NiSe_2$/Ni hybrid foam[65] | 0.136 | 0.5 M $H_2SO_4$ |
| $CoS_2$/graphene/CNT nanocomposite[66] | 0.142 | 0.5 M $H_2SO_4$ |
| Quasi-amorphous $MoS_2$-coated $CoSe_2$ hybrid[67] | 0.068 | 0.5 M $H_2SO_4$ |
| $CoS_2$ nanowires on graphite[68] | 0.145 | 0.5 M $H_2SO_4$ |
| $CoS_2$ nanoparticles/CNT hybrid[69] | 0.061 | 0.5 M $H_2SO_4$ |
| CoSP nanoparticles/CNT hybrid[110] | 0.048 | 0.5 M $H_2SO_4$ |
| CoSP nanoparticles on carbon fibers[70] | 0.048 | 0.5 M $H_2SO_4$ |
| CoMoP nanocrystal coated by a few-layer N-doped C shell[71] | 0.041 | 0.5 M $H_2SO_4$ |
| | 0.081 | 1 M KOH |
| Carbon-Shell-Coated FeP Nanoparticles[72] | 0.071 | 0.5 M $H_2SO_4$ |
| FeP nanoparticles on Ti[73] | ~0.050 | 0.5 M $H_2SO_4$ |
| | 0.102 | Phosphate buffer (pH 7) |
| $Ni_2P$ nanoparticles[74] | 0.115<br>0.130 at 20 mA $cm^{-2}$ | 0.5 M $H_2SO_4$ |
| $NiMoS_3$ Nanorods[75] | 0.212 | 0.5 M $H_2SO_4$ |
| | 0.126 | 1 M KOH |
| $MoS_{2(1-x)}P_x$ solid solution[76] | 0.150 | 0.5 M $H_2SO_4$ |
| WC nanowalls[77] | 0.160 | 0.5 M $H_2SO_4$ |
| CoS film[78] | 0.180 | Phosphate buffer (pH 7) |
| MoP/CNT hybrid[79] | 0.83 | 0.5 M $H_2SO_4$ |
| | 0.102 | Phosphate buffer (pH 7) |
| | 0.86 | 1 M KOH |
| MoP nanoparticles[80] | 0.140<br>~0.180 at 30 mA $cm^{-2}$ | 0.5 M $H_2SO_4$ |
| | 0.130 | 1 M KOH |
| $MoS_2/Mo_2C$ hybrid nanosheets on carbon paper[81] | 0.063 | 0.5 M $H_2SO_4$ |
| Porous $MoC_x$ nano-octahedrons[82] | 0.142 | 0.5 M $H_2SO_4$ |
| | 0.151 | 1 M KOH |
| Co-doped $FeS_2$ nanosheets/CNT hybrids[83] | 0.120 at 20 mA $cm^{-2}$ | 0.5 M $H_2SO_4$ |
| MoB nanoparticles[84] | 0.190 | 0.5 M $H_2SO_4$ |
| | 0.212 | 1 M KOH |
| Ni nanoflakes[85] | 0.08 | 1 M KOH |
| Ni-Mo nanopowders[86] | 0.08 at 20 mA $cm^{-2}$ | 0.5 M $H_2SO_4$ |
| | 0.07 at 20 mA $cm^{-2}$ | 2 M KOH |
| $NiN_3$ on Ni foam[87] | 0.15 at 100 mA $cm^{-2}$ | |
| NiO/Ni heterostructures[88 | 0.08 | 1 M KOH |
| $Ni_{0.89}$-$Co_{0.11}Se_2$ 3D mesoporous nanosheet networks supported on Ni foam[89] | 0.052 | 0.5 M $H_2SO_4$ |
| | 0.082 | Phosphate buffer (pH 7) |
| | 0.085 | 1 M KOH |
| $NiCo_2P_x$ nanowires[90] | 0.104 | 0.5 M $H_2SO_4$ |
| | 0.063 | Phosphate buffer (pH 7) |
| | 0.058 | 1 M KOH |
| C-encapsultated NiCu[91] | 0.048 | 0.5 M $H_2SO_4$ |
| | 0.164 | Phosphate buffer (pH 7) |
| | 0.074 | 1 M KOH |
| $Ni(OH)_2$ decorated $Fe_2P$ nanoarray on Ti mesh[92] | 0.094 | 1 M KOH |
| $Ni_3S_2$ nanowires[93] | 0.199 | 1 M KOH |
| Fe-doped $Ni_3C$ nanodots in N-doped carbon nanosheets[94] | 0.292 | 1 M KOH |
| P-doped $CoS_2$[95] | 0.053 | 0.5 M $H_2SO_4$ |
| $CoS_2$@$MoS_2$/RGO[96] | 0.098 | 0.5 M $H_2SO_4$ |
| CoP nanoparticles[97] | 0.075 | 0.5 M $H_2SO_4$ |
| CoP nanowires on carbon cloth[98] | 0.209 | 1 M KOH |
| N-doped nanoporous C membranes with Co/CoP Janus-type nanocrystals[99] | 0.135 | 0.5 M $H_2SO_4$ |
| | 0.138 | 1 M KOH |
| CoO/Co/N-doped carbon hybrid[100] | 0.232 | 1M KOH |
| Co-embedded N-rich CNTs[101] | 0.380 | 1 M KOH |
| $CoN_x$/carbon catalyst[102] | 0.170 | 1 M KOH |
| Co nanotubes decorated with $TiO_2$ nanodots supported on carbon fibers[103] | 0.106 | 1 M KOH |